\documentclass[twocolumn,times]{aastex631}
\def\NHUNIT{\ifmmode {\rm \,cm^{-2}} \else $\rm \,cm^{-2}$ \fi} 

\def\nhh{\ifmmode N_{\rm H_{2}}\else $N_{\rm H_{2}}$\fi} 
\def\nhhc{\ifmmode N_{\rm H_{2}}^0\else $N_{\rm H_{2}}^0$\fi} 
\def\nhhbg{\ifmmode N_{\rm H_{2}}^{\rm bg}\else $N_{\rm H_{2}}^{\rm bg}$\fi} 
\def\nh{\ifmmode N_{\rm H}\else $N_{\rm H}$\fi}
\def\ml{\ifmmode M_{\rm line}\else $M_{\rm line}$\fi}  
\def\sunpc{\ifmmode \rm M_\odot\,\rm pc^{-1}\else $\rm M_\odot\,\rm pc^{-1}$\fi}  
\def\Msun{\ifmmode \rm M_\odot\else $\rm M_\odot$\fi}  
\def\rflat{\ifmmode R_{\rm flat}\else $R_{\rm flat}$\fi}   
\def\kms{\ifmmode {km\,s$^{-1}$}\else km\,s$^{-1}$\fi}  

\newcommand{\MK}[1]{\textcolor{black}{#1}}
\newcommand{\SA}[1]{\textcolor{black}{#1}}
\newcommand{\DA}[1]{\textcolor{black}{#1}}

\newcommand{\DArr}[1]{\textcolor{black}{#1}}
\newcommand{\rev}[1]{\textcolor{black}{#1}}

\accepted{\today}
\accepted{by ApJL}

\begin{document}

\title{Insights on the Sun birth environment in the context of star-cluster formation in hub-filament systems}

\correspondingauthor{Doris Arzoumanian}
\email{doris.arzoumanian@nao.ac.jp}

\author{Doris Arzoumanian}
\affiliation{National Astronomical Observatory of Japan, Osawa 2-21-1, Mitaka, Tokyo 181-8588, Japan}

\author{Sota Arakawa}
\affiliation{Japan Agency for Marine-Earth Science and Technology, 3173-25, Showa-machi, Kanazawa-ku, Yokohama, 236-0001, Japan}

\author{Masato I.N. Kobayashi}
\affiliation{National Astronomical Observatory of Japan, Osawa 2-21-1, Mitaka, Tokyo 181-8588, Japan}

\author{Kazunari Iwasaki}
\affiliation{National Astronomical Observatory of Japan, Osawa 2-21-1, Mitaka, Tokyo 181-8588, Japan}

\author{Kohei Fukuda}
\affiliation{Forefront Research Center, Osaka University, 1-1 Machikaneyama-cho, Toyonaka, Osaka 560-0043, Japan}
\affiliation{Department of Earth and Space Science, Osaka University, 1-1 Machikaneyama-cho, Toyonaka, Osaka 560-0043, Japan}

\author{Shoji Mori}
\affiliation{Astronomical Institute, Tohoku University, 6-3 Aoba, Aramaki, Aoba-ku, Sendai, Miyagi 980-8578, Japan}

\author{Yutaka Hirai}
\affiliation{Department of Physics and Astronomy, University of Notre Dame, 225 Nieuwland Science Hall, Notre Dame, IN 46556, USA}
\affiliation{Astronomical Institute, Tohoku University, 6-3 Aoba, Aramaki, Aoba-ku, Sendai, Miyagi 980-8578, Japan}

\author{Masanobu Kunitomo}
\affiliation{Department of Physics, Kurume University, 67 Asahimachi, Kurume, Fukuoka 830-0011, Japan}

\author{M. S. Nanda Kumar}
\affiliation{Instituto de Astrof\'isica e Ci{\^e}ncias do Espa\c{c}o, Universidade do Porto, CAUP, Rua das Estrelas, 4150-762 Porto, Portugal}

\author{Eiichiro Kokubo}
\affiliation{National Astronomical Observatory of Japan, Osawa 2-21-1, Mitaka, Tokyo 181-8588, Japan}






\begin{abstract}
 Cylindrical molecular filaments are observed to be  the  main sites of Sun-like star formation, while  massive stars form in dense hubs, at the junction of multiple filaments. The role of 
 hub-filament configurations  has not been discussed yet in relation to the birth environment of the solar system and to infer the origin of isotopic ratios of Short-Lived Radionuclides (SLR, such as $^{26}$Al) of \DArr{Calcium-Aluminum-rich} Inclusions (CAIs) observed in meteorites.
   In this work,  we present simple analytical estimates of the impact of stellar feedback on 
   the young solar system forming along a filament of a hub-filament system. 
   We find  that the host filament can shield  the   young solar system from the stellar feedback, both during the formation and evolution of stars (stellar outflow, wind, and radiation) and at the end of their life (supernovae). We show that  the young solar system formed along a dense filament can be enriched with 
    supernova ejecta (e.g., $^{26}$Al)
   during the formation timescale of CAIs. 
   We also propose that the  streamers recently observed around protostars may be 
  channeling the SLR-rich material %
  onto the young solar system. 
   We conclude that considering  hub-filament configurations as the birth environment of the Sun is important when deriving theoretical models explaining the observed properties of the solar system.
               \end{abstract}



\section{Introduction}\label{intro}

One of the fundamental question related to our existence is that of the origin of our planet Earth and its environment favorable for the emergence of life. This question is also linked to the quest for life on other planets and the issue of how remarkable or unremarkable is our presence in the Universe. The origin of life on Earth is deeply connected to the formation process of the host star.

The solar system is now about 4.6\,Gyr old and the birth environment of the Sun has long since been dissipated. 
Thanks to observational studies describing  the composition of primitive components in meteorites (found on earth or in space), it is possible, in combination with extrapolations and theoretical models, to infer how, when, and the environment where the Sun and its planets were formed \citep[e.g.,][]{Dauphas2011, Nittler2016}. 
The oldest of these primitive condensates are the Calcium-Aluminum-rich Inclusions (CAIs) believed to be 
formed in the protoplanetary disk during the earliest evolutionary time of the solar system \citep[e.g.,][]{Amelin2002}.  
Isotopic studies of CAIs found in meteorites show the presence of several Short-Lived Radionuclides (SLRs) such as the $^{26}$Al, which provide constraints on the formation timescales and processes of the CAIs and on the amount and timescale of newly injected material as a result of feedback from neighbouring high-mass stars \citep[e.g.,][]{Scott2007,Adams2010,Boss2012}. %
To understand the origin of these observations of the solar system properties it is essential to characterise the molecular cloud structure where the Sun formed and understand the impact of feedback from neighboring stars on the young solar system.

Star formation has traditionally been presented  as taking place in  spherical 3D clouds of dust and mostly molecular gas \citep[e.g.,][]{Bergin2007}. These clouds fragment into prestellar cores, the seeds of future stars \citep[e.g.,][]{Ward-Thompson1994,Ward-Thompson1999}. %
These gravitationally bound prestellar cores collapse into a stellar embryo, a protostar, which evolve through accretion (going through different phases from Class 0 to III) reaching the pre-main sequence phase with a star surrounded by a circumstellar disk \citep[e.g.,][]{Andre2000}.

Recent star formation studies described the detailed density distribution of star forming molecular clouds
\citep[e.g.,][for a review]{Andre2014}. 
It is now widely accepted that 
molecular clouds are filamentary, and that dense molecular filaments are the main sites of low- to intermediate-mass star formation \citep[see the recent review by][]{Pineda2022}, while 
high-mass stars are preferentially formed in hubs formed by the junctions of multiple filaments \citep{Kumar2020,Kumar2022}. In this star formation context, filaments are elongated (high aspect ratio) cylindrical molecular gas structures and hubs are high density (low aspect ratio) compact  clump-like structures where two or more filaments merge \citep{Myers2009}.

To study the formation and evolution of stars and their planetary systems it is thus crucial to take into account the dense, non-uniform, and filamentary strutcure of the host molecular cloud. In particular the impact of stellar feedback  on the surrounding cloud and young stars might be highly affected by these hub-filament configurations where star-clusters are observed to be forming.
The aim of this paper is thus to discuss the role of  hub-filament systems in the understanding of the origin of the properties of Sun-like stars and their planetary systems,  in particular our solar system.
To do that we first give a brief overview on the observational constraints of the abundance of SLRs in CAIs in the solar system  and the theoretical understanding of their formation processes and sites  (Section\,\ref{sec:SLR}). We then describe the current understanding of the star formation process (Section\,\ref{sec:HFS}). By combining the understanding presented in the previous sections, we  discuss the impact of feedback from massive stars on Sun-like stars forming along the filaments   (Section\,\ref{feedback}) and the role of the hub-filament configuration in shielding the protosolar system from the destructive effect of stellar feedback while providing the required amount of newly injected material onto the protosolar system that would explain the observations (Section\,\ref{Discussion}). We summarize and conclude the paper in Section\,\ref{conclusion}.

\section{Overview of SLR studies and open questions}\label{sec:SLR}

\subsection{Formation sites and processes of SLRs}\label{sec:SLR1}

In this study, we define SLRs as radioactive nuclides that were present in the early solar system but now extinguished.
Their existence is recorded in meteorites as excesses of the daughter isotopes.
There are three types of sources for the SLRs in the early solar system: (i) SLRs in the parental molecular cloud (i.e., ``inheritance''), (ii) production of SLRs via irradiation of high-energy cosmic rays around the young Sun, and (iii) injection of SLR-rich materials from nearby massive stars, e.g., Wolf-Rayet (WR) winds and core-collapse supernovae.
\DArr{Below, we briefly summarize the current understanding of the three type of sources, which in combination are required to explain the  observed SLR content of the early solar system.}

\subsubsection{SLRs enrichment in the parental molecular cloud}

The first source for SLRs in the early solar system is the background level before its formation.
Two terms for the background level of SLRs are usually considered: the galactic background and the self-enrichment within the parental molecular cloud.
The galactic background is the amount of SLRs in the interstellar medium that forms the parental molecular cloud.
The contribution of the galactic background for the amount of SLRs in the early solar system can be estimated from a simple theoretical model called the galactic chemical evolution model \citep[see Section 2 of][]{Huss2009}.
\SA{We note, however, that the initial abundance of $^{26}{\rm Al}$ in the solar system is orders of magnitude higher than that predicted by the galactic background \citep[e.g.,][]{Huss2009}.}

To solve this discrepancy, several scenarios have been proposed.
The self-enrichment within the parental molecular cloud by sequential star formation and death of massive stars would provide a considerable amount of SLRs \citep[e.g.,][]{Gounelle2012}.
\citet{Fujimoto2018} also proposed an alternative scenario that SLRs in giant molecular clouds originate from ejecta of massive stars that will subsequently be incorporated into existing or newborn clouds \citep[see also][]{Young2014}.

Although these scenarios proposing that SLRs originated from sequential star formation \citep[e.g.,][]{Gounelle2012, Fujimoto2018} can explain the abundance of SLRs recorded in normal CAIs, these scenarios would have difficulty in reproducing the coexistence of $^{26}{\rm Al}$-poor unusual CAIs and $^{26}{\rm Al}$-rich normal CAIs (see Section \ref{sec.FUN}).

\subsubsection{Irradiation of high-energy cosmic rays.}

The second source for SLRs in the early solar system is the production of SLRs via irradiation of high-energy cosmic rays.
When the cosmic ray energy is sufficiently large \citep[e.g., $\gtrsim 10^{7}~{\rm eV}$ for production of $^{26}{\rm Al}$;][]{Gaches2020}, interaction with atomic nuclei produces smaller fragments including some SLRs.
It is widely accepted that $^{10}{\rm Be}$ was produced via irradiation of high-energy cosmic rays because it cannot be synthesized by thermonuclear reactions inside massive stars \citep[e.g.,][]{Lugaro2018}.
There are two possible sources for high-energy cosmic rays: solar cosmic ray irradiation in the circumstellar disk \citep[e.g.,][]{Jacquet2019, Fukuda2021} and galactic cosmic ray irradiation possibly in the parental molecular cloud \citep[e.g.,][]{Desch2004, Dunham2022}.

Recently, \citet{Gaches2020} proposed a local mechanism to enrich $^{26}{\rm Al}$ via solar cosmic ray irradiation.
We note, however, that this model requires a low mass accretion rate \DArr{through} the circumstellar disk to reproduce the initial ratio of ${( ^{26}{\rm Al} / ^{27}{\rm Al} )}$, while \DArr{high mass accretion rates are often observed \citep[e.g.,][]{Hartmann2016} during the expected formation epoch of CAI.}

\subsubsection{Direct injection of SLR-rich materials to the Sun's prestellar core/protoplanetary disk}\label{sec:SLR13}

The third source for SLRs in the early solar system is the direct injection of SLR-rich materials from nearby evolved massive stars.
In active star-forming regions, several massive stars could be formed.
As the lifetime of massive stars is short ($< 10~{\rm Myr}$ for stars with a mass  larger than 20\, M$_{\sun}$), direct injection from nearby evolved massive stars into newborn Sun-like stars would be possible in the parental molecular cloud.
There are two candidates for the origin of SLR-rich materials: Wolf-Rayet (WR) winds and core-collapse supernovae (SNe).

It is thought that the initial abundance of $^{60}{\rm Fe}$ is the key to unveiling the source of SLRs.
This is because $^{60}{\rm Fe}$ is barely produced by energetic-particle irradiation around the young Sun \citep[e.g.,][]{Lee1998}, and the abundance ratio of $^{60}{\rm Fe}$ and $^{26}{\rm Al}$ significantly differs among stellar sources.
Recent measurements of the initial $^{60}{\rm Fe}$ abundances in chondrules and chondritic troilites support the lower initial abundances of $^{60}{\rm Fe}$ in the early solar system \DArr{\citep[$^{60}{\rm Fe} / ^{56}{\rm Fe} \lesssim 10^{-8}$;][]{Tang2015,Trappitsch2018,Kodolanyi2022,Kodolanyi2022New},} although several studies reported higher initial abundances \citep[$^{60}{\rm Fe} / ^{56}{\rm Fe} \gtrsim 10^{-7}$; e.g.,][]{Mishra2016, Telus2018}.

If the initial $^{60}{\rm Fe} / ^{56}{\rm Fe}$ ratio is $^{60}{\rm Fe} / ^{56}{\rm Fe} \sim 10^{-8}$ or less, no \DArr{supernova} injection of $^{60}{\rm Fe}$-rich material is required 
\DArr{\citep[e.g.,][]{Huss2009}.} %
In this case, WR winds are  one of the leading candidates for the source of SLRs as they are enriched in $^{26}{\rm Al}$ but depleted in $^{60}{\rm Fe}$ \citep[e.g.,][]{Dwarkadas2017}.
SNe with fallback could also explain the low abundance of $^{60}{\rm Fe}$ \citep[e.g.,][]{Takigawa2008, Huss2009}.
\DArr{We note, however, the presence of multiple uncertainties in the production of $^{60}{\rm Fe}$ in core-collapse SN models that depend on, e.g., the uncertain $^{59}{\rm Fe}$$^{60}{\rm Fe}$ cross section and the assumed reaction rate  \citep[e.g.,][]{Jones2019}.}

The progenitor mass of a plausible SN (or a WR star) that affected the early solar system is still poorly constrained.
\citet{Huss2009} proposed that a SN with mixing-fallback could reproduce the initial abundances of SLRs when the progenitor mass is in the range of $20$--$60\,\Msun$ \citep[see also][]{Takigawa2008}.
However, their mass estimate is based on the assumption of higher initial $^{60}{\rm Fe} / ^{56}{\rm Fe}$ ratio in classical literature \citep[$^{60}{\rm Fe} / ^{56}{\rm Fe} \gtrsim 10^{-7}$;][]{Tachibana2003}, and the mass range for the plausible progenitor should be updated.
\citet{Sieverding2020} proposed that a low-mass progenitor ($11.8\,\Msun$) of the SN would also be suitable to resolve the overproduction problem of $^{60}{\rm Fe}$.
The minimum stellar mass for evolving into WR stars is approximately $30\,\Msun$, although that depends on the stellar metallicity and the rotation velocity \citep[e.g.,][]{Georgy2012}.

The timing of injection is also under debate.
One of the plausible scenario is the direct injection onto the prestellar core \citep[e.g.,][]{Cameron1977, Gritschneder2012}, and another scenario is injection to the circumstellar disk \citep[e.g.,][]{Clayton1977, Ouellette2010}.
\citet{Fukai2021} argued that injection of SN materials onto the circumstellar disk could be reasonable when the variation of stable Cr isotope ratio recorded in bulk carbonaceous chondrites are inherited from the disk-scale spatial heterogeneity.
Note that we cannot rule out the other scenario that direct injection onto the prestellar core caused the spatial heterogeneity recorded in Cr isotope ratio.

\subsection{CAIs, the first condensates formed within the early solar system}
\label{sec.FUN}

CAIs are the oldest dust particles condensed in the solar protoplanetary disk 4.567 billion years ago \DArr{\citep[e.g.,][]{Amelin2002,Connelly2012}}.
The presence of $^{26}{\rm Al}$ in the early solar system is supported by the excess of its daughter product, $^{26}{\rm Mg}$, in meteorites.
Both ${\rm Al}$ and ${\rm Mg}$ are abundant in CAIs, and their isotopic abundances can be determined precisely.
The initial ratio of $^{26}{\rm Al} / ^{27}{\rm Al}$ at the timing of CAI formation, ${( ^{26}{\rm Al} / ^{27}{\rm Al} )}_{\rm initial}$, is approximately $5 \times 10^{-5}$ \citep[e.g.,][]{MacPherson2012}.

Most of the primitive (i.e., unmelted) CAIs are enriched in $^{26}{\rm Al}$ \citep[e.g.,][]{MacPherson2012}; however, some unusual CAIs show a very low initial abundance of $^{26}{\rm Al}$.
One of those unusual CAIs are platy hibonite crystals (PLACs), and they are regarded as the oldest CAIs because of their high condensation temperature and large nucleosynthetic anomalies \citep[e.g.,][]{Koeoep2016}.
CAIs with fractionation and unidentified nuclear effects (FUN CAIs) have also low and varied initial abundance of $^{26}{\rm Al}$, thought to be formed prior to normal CAIs \citep[e.g.,][]{Park2017}.
The coexistence of $^{26}{\rm Al}$-rich (normal) CAIs and $^{26}{\rm Al}$-poor CAIs (PLACs and FUN CAIs) is regarded as the evidence of the direct injection of $^{26}{\rm Al}$-rich materials onto the early solar system at the epoch of CAI formation \citep[e.g.,][]{Sahijpal1998, Holst2013}.
Since the duration of CAI formation is a few $10^{5}$ years or less \citep[e.g.,][]{Connelly2012}, the injection event should have occurred in the first few $10^{5}$ years of the solar system formation \DArr{\citep[][]{Arakawa2022}.}

Although normal CAIs are uniformly enriched in $^{26}{\rm Al}$, the abundance of $^{26}{\rm Al}$ in FUN CAIs is significantly varied.
\citet{Park2017} reported that most FUN CAIs show the initial isotopic ratio for individual CAIs, ${( ^{26}{\rm Al} / ^{27}{\rm Al} )}_{0}$, in the range from $5 \times 10^{-8}$ to $5 \times 10^{-5}$.
The abundance of $^{26}{\rm Al}$ in PLACs is significantly low; ${( ^{26}{\rm Al} / ^{27}{\rm Al} )}_{0} \ll 1 \times 10^{-5}$ and it is frequently undetectable \citep[e.g.,][]{Koeoep2016}.
We can interpret these \DA{observed variations of} $^{26}{\rm Al}$ abundances as the consequence of \DA{the timescale and the} injection \DA{processes} of $^{26}{\rm Al}$-rich materials and CAI formation.
PLACs might be formed prior to the injection of $^{26}{\rm Al}$-rich materials, FUN CAIs would be formed during the injection and mixing of solar and extrasolar materials, and normal CAIs are formed after injection and mixing \rev{\citep[cf.][]{Krot2019}}.

\subsection{Open questions}

As presented in Sections\,\ref{sec:SLR1} and \ref{sec.FUN}, because of the coexistence of $^{26}{\rm Al}$-rich and $^{26}{\rm Al}$-poor CAIs, 
we favor the %
injection of SLR-rich materials  through a SN event/WR wind from a nearby evolved massive star. 

Several studies, however, pointed out the difficulty of such a scenario due to the survivability of the pre-solar system to the impact of the SN shock. 
For example, \citet{Kinoshita2021} found that nearby SNe would disrupt the molecular prestellar cores instead of triggering the collapse of these cores into protostars \citep[as suggested by, e.g.,][]{Gritschneder2012} when the shock is too strong.
At a more evolved stage, \citet{Close2017} show that a circumstellar disk might be disrupted unless the disk is close to edge-on with respect to the SN expansion direction, while the cross section for enrichment of SLR-rich materials would be too small to reproduce the initial value of the solar system when the disk is close to edge-on.

All previous works, however, study the impact of stellar feedback on a pre-Sun (prestellar core stage) or proto-Sun (with a circumstellar disk)  surrounded by a low density ($\sim10^2$\,cm$^{-3}$) environment.
Recent observations have shown that the bulk of star formation takes place along dense  molecular filaments (see Section\,\ref{sec:HFS}), which could play a significant role in shielding the star forming core from the stellar feedback and modifying the supply process of SLR-rich materials to the pre-solar system.
Below, we elaborate on the link between the filamentary structure of molecular clouds and the impact of the environment on the properties of the solar system in formation.

\vspace{2cm}
\section{Overview of the current understanding of the star formation process}
\label{sec:HFS}

\subsection{Star formation from molecular filament fragmentation}\label{sec:HFS1}

For a long time,  the Galactic interstellar medium (ISM) has been known to be filamentary \citep[][]{Schneider1979}.
It is only recently, however, that the ubiquity of filaments in the cold ISM have been revealed 
\citep[e.g.,][]{Andre2010,Molinari2010}. 
In molecular clouds, the detailed analyses of the radial density profile of molecular filaments 
show that filaments are characterized by Plummer-like density distribution, with a flat inner width of $\sim$0.1\,pc and a power-law density distribution at large radii with a power-law exponent $\sim$2 \citep[][]{Arzoumanian2011,Arzoumanian2019,Palmeirim2013,Andre2022}. These filaments span a wide range in central column density and mass per unit length \rev{($M_{\rm line}$). Filaments with $M_{\rm line}>M_{\rm line,crit}$ are thermally supercritical and unstable against gravitational fragmentation and star formation. Here,} $M_{\rm line,crit}=2 c_{\rm s}^{2}/G \sim16\,\sunpc$
is the critical value of isothermal unmagnatized cylinders  \citep[cf.][]{Inutsuka1997},
where $c_{\rm s} \sim0.2$\,km\,s$^{-1}$  is the isothermal sound speed for cold molecular gas at $T_{\rm gas} \sim 10$\,K.
Star forming supercritical filaments are also observed to be in rough virial balance with 
$M_{\rm line}\sim M_{\rm line,vir}>M_{\rm line,crit}$ \citep[][]{Arzoumanian2013} and
 \begin{equation} \label{eq:Mline}
M_{\rm line}=\Sigma_{\rm fil}\,W_{\rm fil},
\end{equation}
where $\Sigma_{\rm fil}$ and $W_{\rm fil}$ are the filament surface density and width, respectively, and
  \begin{equation} \label{eq:Mvir}
M_{\rm line,vir}=2\sigma^2_{\rm tot}/G,
\end{equation}
where $\sigma_{\rm tot}$ is the total (thermal + turbulent) velocity dispersion. 

A large fraction of prestellar cores and protostars are indeed observed along supercritical virialzed filaments and the  mass distribution of these cores, the prestellar core mass function (CMF), has a shape similar to the initial mass function (IMF) of stars  shifted by $\sim30\%$ to larger masses \citep[e.g.,][]{Andre2010,Konyves2015,Konyves2020}.
Moreover, recent studies proposed that the 
typical mass of the cores formed along a given filament is set by the filament line mass \citep{Andre2019,Shimajiri2019ALMA} and it corresponds  to 
the critical Bonnor-Ebert mass \citep{Bonnor1956} at the local filament surface density and effective temperature, with
  \begin{equation} \label{eq:MBE}
M_{\rm BE,eff}  \sim \frac{1.3\,\sigma_{\rm tot}^4}{G^2\,\Sigma_{\rm fil}}.  
\end{equation}
A detailed analysis of the filament properties has shown the wide range of filament mass per unit length  suggesting that 
the  shape of the CMF, and by extension that of the stellar IMF, may be partly inherited from the filament line mass function (FLMF) where higher-mass cores form in higher-line mass filaments \citep[][]{Andre2019,Pineda2022}. 
For example, a $1\,\Msun$ sun-like star would result from the collapse and evolution of a 
$M_{\rm pre}\sim3\,\Msun$ 
presetellar core (for an efficiency of $\sim30\%$). Such a 
core would form along a 0.1\,pc-wide 
supercritical filament with a 
line mass of

 \begin{equation}\label{eq:MlineSun}
 M_{\rm line}^{{\rm fil-Sun}}\sim90\,\sunpc\left(\frac{M_{\rm pre}}{3\,{\rm M}_{\odot}}\right)\left(\frac{W_{\rm fil}}{0.1\,{\rm pc}}\right)^{-1}
  \end{equation}
derived by combining Eqs.\,\ref{eq:Mline}, \ref{eq:Mvir} and \ref{eq:MBE}.  
This filament line mass corresponds to a peak column density of $\sim4\times10^{22}\,\NHUNIT$ and a density of 
\DArr{$\Sigma_{\rm fil}/W_{\rm fil}$}$\,\sim10^{5}\,$cm$^{-3}$ 
(assuming the filament on the plane-of-the-sky \DArr{and the line-of-sight length is equal to the filament width $W_{\rm fil}$}). 

\subsection{Role of hub-filament systems in star-cluster formation}\label{sec:HFS2}

Molecular filaments are not isolated objects, but are observed to form networks 
with multiple junctions. 
The intersection between two or more filaments are referred to as hubs \citep[cf.][]{Myers2009}. The system of hubs and the surrounding filaments are usually termed as a hub-filament system (HFS). 
These HFSs have been suggested to be particularly favorable for the formation of star clusters and massive stars thanks to inflows of matter along the filaments, traced by the observed velocity gradients, that increase the mass of the hubs where massive stars can form  \citep[e.g,][]{Peretto2013,Peretto2014,Williams2018,Chen2019}. 
A statistical study of the environment of the massive luminous sources in the  Milky-Way shows that all high-mass stars preferentially form in the dense hubs of hub-filament systems \citep{Kumar2020}.
A recent detailed analysis of the structure of the hub in the Mon R2  HFS reveals a new view of the hub as a network of high-density filaments, as opposed to its previous description of massive clump \citep[][]{Kumar2022}. In these hubs, only stars located in the filament network can benefit from  longitudinal flows of gas to become massive, which may explain the reason for the formation of many low-mass stars in cluster centres, and the possibility for the young stellar systems  to be effectively shielded from the  (destructive) feedback of nearby massive stars.

Thus, the HFS configuration can explain  the following features of star cluster formation \citep[cf.][]{Kumar2020}: (a) low- and intermediate-mass stars form first and slowly ($\sim10^6$\,yr) along the filaments, (b) massive stars form later, after the formation of the hub, and their formation is faster than the lower mass stars  ($\sim10^5$\,yr), 
 (c) the massive star feedback is dissipated through the inter-filamentary medium,  
(d) the initial mass function of stars is the combination of the stars continuously formed along the filaments with all massive stars formed in the hub resulting in (e) mass segregated clusters and (f) an age spread of the stars within bound clusters. 

\begin{figure*}[!h]
 \centering
   \resizebox{18.cm}{!}{\includegraphics[angle=0]{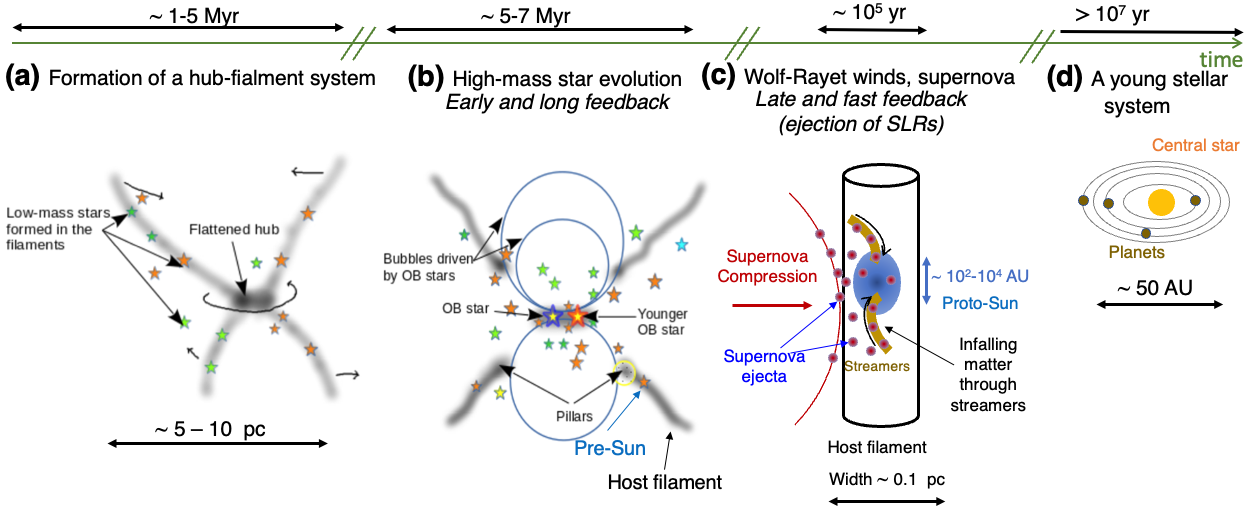}}
\caption{Schematic figure summarizing the importance of hub-filament systems (HFS) in the understanding of the impact of stellar feedback on the surrounding matter and the formation of stellar systems. (a) Formation of a HFS through the merging of multiple filaments forming low- to intermediate-mass stars (Sections\,\ref{sec:HFS1} and\,\ref{sec:HFS2}). (b) High-mass star formation in the dense hub. The feedback from massive stars during their formation and evolution (outflow and wind) compresses the dense filaments \citep[see][]{Zavagno2020} and flow away into the ISM through the inter-filamentary diffuse medium forming bipolar shaped  HII  regions (Section\,\ref{HII}). (c) At the end of their lives high-mass stars go through a Wolf-Rayet phase followed, for some of them, by a supernova explosion (Section\,\ref{SNe}). During these later stages of high-mass star evolution, SLRs (such as $^{26}$Al) present in the Wolf-Rayet winds and supernova ejecta (Section\,\ref{sec:SLR}) are injected into the filament and the cores embedded in them. These cores are additionally supplied with SLR rich gas and dust by streamers throughout the different protostellar evolutionary phases (Section\,\ref{sec:streamers}) until the formation of planetary systems (d). (a) and (b) are adapted from \citet{Kumar2020}.}
  \label{SummaryFig}
  \end{figure*}

\subsection{Streamers and accretion from core to disk}
\label{sec:streamers}

 Another important consideration in the formation of star and planetary  systems is the assembly, origin, and composition of the dense gas material that will eventually form the star and the protoplanetary disk. 
 As mentioned above, the first step of the formation of solar type stars has been identified as a starless  phase, where a gravitationally bound prestellar core \citep{Ward-Thompson1994,Alves2001,Caselli2002} forms from filament fragmentation. 
 Such prestellar cores  \rev{were believed to encompass  the matter reservoir out of which 
a single star or a small stellar system  forms  \citep[with a $\sim30\%$  efficiency for the mass conversion, cf. Sect.\,\ref{sec:HFS1},][]{Andre2014}.}  
 However, recently, this picture is being challenged with new observations showing streamer-like elongated structures connected to protostellar systems, while extending outside of the core envelope  scale \citep[$>0.1\,$pc, see a recent review by][]{Pineda2022}.  Velocity gradients are observed along these streamers compatible with models of free fall motions, 
 suggesting matter flows feeding the young star and its disk. Mean infall rates  of $\sim10^{-6}\,$M$_\odot$\,yr$^{-1}$ per streamer have been inferred from molecular line observations  \citep{Pineda2020,Thieme2022,Valdivia-Mena2022}. 
 In addition, these streamers are observed to be composed of carbon rich species, which are expected to be depleted onto dust grains in prestellar cores. The detection of such molecules in the streamers suggest the replenishment of the  cores with chemically fresh gas and dust from outside of the core envelope. Such streamers are observed towards protostars at different evolutionary stages from Class 0 to Class III, suggesting the long duration of matter infall that may change the final mass of the system as well as its composition.

Such streamer-like structures are also produced in numerical simulations \citep{Kuffmeier2020,Kuffmeier2021} and suggested to result from the capture of cloudlets and their stretching due to the gravity of the protostar \citep[][]{Hanawa2022}. 
While the formation and evolution process of streamers are yet to be studied,
 these recent results raise new questions 
 regarding 
 the origin of the material available to form the star, its protoplanetary disk, and planets, as well as the timescales of the different evolutionary stages of stellar system formation. 
 
\section{Impact of stellar feedback on planetary system formation}\label{feedback}

Stellar feedback from massive stars ($M_\star>8$\,M$_{\odot}$) is believed to have an important impact on the properties of the surrounding ISM contributing to its chemical enrichment in new elements and the injection of momentum and ionising photons. Stellar feedback can be divided into ``early and long lasting" (during the formation and evolution of the star: HII  regions from stellar wind, radiation, and outflows) and ``late and fast" (Wolf-Rayet winds and supernovae events) phenomena (see Figure\,\ref{SummaryFig}). In the following we discuss the impact of these two types of feedback on sun-like star formation, especially we suggest the survival of hub-filament systems to the impact of HII  regions (Section\,\ref{HII}) and the role of the filaments in shielding the forming planetary system from the impact of  supernova shock compression (Section\,\ref{SNe}).

\subsection{HII  regions}\label{HII}

HII  regions are ionized hot bubbles generated by stellar winds, ionizing radiation, and outflows of massive stars. 
Theoretical works have studied the effect of expanding  HII  regions (stellar winds, photoionization, and photodissociation) on the ambient molecular cloud and on the quenching of star formation \DArr{\citep[e.g.,][]{Hosokawa2006,Walch2012,Dale2013,Krumholz2014,JGKim2018}. 
For example,} for a spherical non-magnetized molecular gas of density $\sim10^2$\,cm$^{-3}$ around a massive star of $\gtrsim20$\,M$_{\odot}$ about $\sim3\times10^4$\,M$_{\odot}$ is photodissociated  within 5\,Myr \citep[][]{Inutsuka2015}. 
These calculations, however, do not include %
the non-spherical hub-filament like geometry of the molecular cloud around the massive star (cf., Section\,\ref{sec:HFS2} and Figure\,\ref{SummaryFig}). These dense filamentary structures should significantly affect the actual expansion of the HII  region  as well as its impact on the surrounding cloud and its star formation process \citep[e.g.,][]{Arthur2011}.  

Observations  towards HII   regions often show dense  material inside (pillar-like structures pointing towards the ionizing stars) and at the edge of the bubbles, at distances of  $\sim0.2-0.5$\,pc of the bubble center \citep[][]{Kumar2020}.
There are  examples of bipolar HII  regions \citep[e.g.,][]{Deharveng2015,Samal2018} possibly associated to evolved hub-filament systems where  the ionizing pressure and radiation from the high-mass stars formed in the hub escape through the inter-filamentary gaps, while the dense filaments continue forming stars \citep[][]{Kumar2020,Kumar2022}. 
These observations suggest the high survival rate of dense structures within and at the edge of expanding HII  regions.
Therefore future theoretical studies should investigate the effect of HII  regions  in more realistic three dimensional magnetohydrodynamic simulations to assess the quenching of star formation due to early stellar feedback. 

In the following we thus assume that star-forming filaments survive  the early stellar feedback from massive stars formed in  hub-filament systems and consider the pre-sun to be embedded in a filament  with a line mass of $\sim90\,\sunpc$ 
(see Section\,\ref{sec:HFS1})
and the filament to be  tangential to the expanding shell (Figure\,\ref{SummaryFig}).

\subsection{Supernovae}\label{SNe}

At the end of their lives, after the generation of the HII  regions, high-mass stars will impact their surroundings with their WR winds and  supernova explosions.
In the context of the origin of the SLRs in the solar system,  both WR winds and  supernova events are relevant, with the latter having the most destructive impact. \DArr{We note, however, that not all massive stars explode in supernova, but some become black holes \citep[see, e.g.,][]{Sukhbold2016}.}

As discussed in the previous sections, instead of the direct injection onto the Sun's prestellar core / portoplanetary disk, %
the SLR-rich material may be injected
onto the filament that hosts the proto-Sun
and 
subsequently accreted, e.g.,    
along streamers (cf., Section\,\ref{sec:streamers} and Figure\,\ref{SummaryFig}c), 
onto the protostellar system.
Such  transfer of matter 
may be at the origin of the observed
isotope variation and explain the coexistence of normal CAI and FUN CAI/PLAC.
The key questions are (1) whether the filament can survive during the supernova feedback shielding the embedded solar nebula and (2) 
whether the required amount of $^{26}$Al is channeled onto the solar nebula
to explain the observations.

As a first step, one can estimate the typical destruction timescale of a single filament
experiencing a blast wave of a supernova at the right angle (see Figure\,\ref{SummaryFig}c).
\cite{Klein1994} analytically investigate the interaction between a blast wave and an 
overdense cloud to estimate the typical cloud destruction timescale due to the Kelvin-Helmholtz instability and Rayleigh-Taylor instability.
They also perform two-dimensional adiabatic and isothermal simulations
and derive the typical cloud 
destruction timescale  %
as \citep{Klein1994}:
\begin{equation}
    t_{\rm cc} = \frac{\chi^{1/2} a_0}{v_{\rm b}},
    \label{eq:KMC1994}
\end{equation}
where $\chi=n_{\rm cloud}/n_{\rm intercloud}$ is the density ratio of the cloud and the inter-cloud medium, 
$a_0$  is the radius of a spherical cloud, 
and $v_{\rm b}$ is the blast wave velocity in the cloud rest-frame.

\cite{Nakamura2006} extended  this study by performing two-dimensional and three-dimensional hydrodynamics simulations to investigate the destruction of  spherical and  cylindrical clouds with  power-law density profiles smoothly connected to the inter-cloud medium as
\begin{equation}
    n_{\rm cloud}(r) = n_{\rm intercloud} + \frac{n_{\rm cloud,0}-n_{\rm intercloud}}{1+(r/a_0)^p}.
\end{equation}
Here $n_{\rm cloud,0}$ is the density at the cloud center and $n_{\rm intercloud}$ is the density of the cloud at $r>>a_0$, with 
$r$  the radial distance. %
They varied the density profile index $p$ and  $\chi$ 
and found that the cloud destruction timescale is more affected by $p$ rather than $\chi$ or the geometry of the cloud (either spherical or cylindrical).
For  $p=2$, which corresponds to the observed Plummer-like filament density profile (see Section\,\ref{sec:HFS1}),
\begin{equation}
    t_{\rm dest} \simeq 12\, t_{\rm cc}.
    \label{eq:N2006}
\end{equation}

We hereafter replace $n_{\rm cloud,0}$ by $n_{\rm fil,0}$,
$n_{\rm intercloud}$ by  $n_{\rm interfil}$,
and $a_0$ 
by the filament inner radius $R_{\rm fil} =W_{\rm fil}/2$.
As shown in Section~\ref{sec:HFS1}, we employ
$n_{\rm fil,0}\simeq 10^5$ cm$^{-3}$ 
as the typical volume density of a filament that would form sun-like stars and 
$R_{\rm fil} = 0.05$\,pc as the typical filament radius. 
In a hub-filament configuration, \MK{the medium surrounding such a filament is presumably a part of an H{\sc ii} region
created by the feedback from the pre-supernova stage massive stars in the hub (see Fig.\,\ref{SummaryFig}b), thus 
$n_{\rm interfil} \sim n_{\rm H{\sc ii}} \simeq 10$\,cm$^{-3}$}
\citep{Kennicutt1984}.
Consequently $\chi\sim10^4$.
\MK{Given}
\DA{an explosion energy of $10^{51}$\,erg and $v_{\rm b} \sim 200$ km s$^{-1}$ at the end of the Sedov-Taylor phase \citep{Thornton1998,Kim2015} 
for $n_{\rm interfil} \simeq 10$\,cm$^{-3}$,
}
we can  estimate the filament destruction timescale $t_{\rm dest,fil}$
by combining Eqs.~\ref{eq:KMC1994} and~\ref{eq:N2006}
as

\begin{equation}    \label{eq:tcc}
    t_{\rm dest,fil}\sim 0.3\, \mathrm{Myr}\,
    \left(\frac{\chi}{10^4}\right)^{1/2}
    \left(\frac{R_{\rm fil}}{0.05\,\mathrm{pc}}\right)
\left(\frac{v_{\rm b}}{200\,\mathrm{km\,s^{-1}}}\right)^{-1}. 
\end{equation}
Eq.\,\ref{eq:tcc} gives an estimate of the filament destruction timescale when the shock compression is orthogonal to the filament long axis. Other configurations of the filament orientation will result in larger $t_{\rm dest,fil}$ values.
In addition, hub-filament systems are observed to be magnetized. Magnetic fields play a role in preventing the growth of the \DArr{Kelvin-Helmholtz} instability reducing the destructive impact of the shock on the filament.
Taking into account radiative cooling in the dense post-shock filament during the shock compression would also contribute in increasing the destruction timescale. 
We thus suggest that $t_{\rm dest,fil}\sim 0.3\,$Myr is an estimate of the minimum filament survival time and that  $t_{\rm dest,fil}$ would be larger if other processes (e.g., magnetic field) and other configurations %
are considered. 
Thus a star forming filament may survive a SN blast for $>0.3\,$Myr while the SN ejecta is injected in the filament and onto the star forming systems (Figure\,\ref{SummaryFig}c).

\section{SLR-rich material supply onto the young solar system embedded in a molecular  filament}\label{Discussion}

In Section\,\ref{feedback} we described the important role that the molecular filament plays in protecting the embedded solar protoplanetary disk from the destructive \rev{feedback} effect from \rev{nearby massive stars.}
Here we evaluate whether a solar system embedded in a filament may receive the required amount of ejecta from \rev{the feedback from a nearby massive star.}
\rev{To do that we first estimate} the initial mass of $^{26}$Al in the solar nebula at the time of normal CAI formation.
For simplicity, we consider the situation where the $^{26}$Al-rich materials 
are initially \rev{injected and} mixed in the filament that host the proto-Sun system, and subsequently channeled to and mixed in the solar protoplanetary disk.
\rev{Second,  the amount of SLRs that reach the early solar system needs to be estimated. This latter amount, however, is uncertain, highly debated, and depends strongly on the mass of the progenitor,  the stellar evolution models (i.e., including or not magnetic fields and rotation),  the nucleosynthesis  models, and whether or not the massive star explodes in an SN \citep[e.g.,][]{Sukhbold2016}. In the following, we  consider  the case where the source of $^{26}$Al is a  $25\,\Msun$ progenitor that explodes in an SN\footnote{\rev{We note that according to some models not all  stars in the $22$--$25\,\Msun$ range explode in SNe \citep[e.g.,][]{Sukhbold2016}.}}. 
We then show that a similar amount of $^{26}$Al can be produced by the WR winds of a $60\,\Msun$ star that does not explode in an SN. A more extensive parameter study of progenitor masses and origin of SLRs will be presented in a follow-up work.}

\subsection{Initial mass of $^{26}$Al in the solar protoplanetary disk at the timing of normal CAI formation}\label{IniMass}

The initial mass of $^{26}$Al in the protoplanetary disk, $M_{\rm 26Al}^{\rm PPD}$, is given by the following equation:
\begin{equation}
M_{\rm 26Al}^{\rm PPD} \simeq  M_{\rm PPD} \, {( M_{\rm Al} / M_{\rm H} )}_{\rm solar}\,  {( M_{\rm 26Al} / M_{\rm Al} )}_{\rm initial},
\label{IniMassPPD}
\end{equation}
where $M_{\rm PPD}$ is the mass of the solar protoplanetary disk,
${( M_{\rm Al} / M_{\rm H} )}_{\rm solar}$ is the solar abundance of $^{27}$Al, and  ${( M_{\rm 26Al} / M_{\rm Al} )}_{\rm initial}$ is the isotopic ratio of $^{26}$Al at the timing of normal CAI formation.  
Assuming that $M_{\rm PPD} \sim 10^{-1} M_{\sun}$, $\log{( M_{\rm Al} / M_{\rm H} )}_{\rm solar} = - 5.57$ \citep{Asplund2021}, and ${( M_{\rm 26Al} / M_{\rm Al} )}_{\rm initial} = 5 \times 10^{-5}$ \citep{MacPherson2012}, we can perform an order-of-magnitude estimate for $M_{\rm 26Al}^{\rm PPD}$, and we obtain
$M_{\rm 26Al}^{\rm PPD} \sim 10^{-11} M_{\sun}$.

\rev{We here assume  $M_{\rm PPD} \sim 10^{-1} M_{\sun}$, which  corresponds to the theoretically estimated total mass of the solar nebula at the end of the class 0 stage \citep[cf. e.g.,][for a recent review]{Tsukamoto2022}. We acknowledge, however, the large uncertainty on this assumed value and foresee future parameter studies exploring possible ranges of $M_{\rm PPD}$ for different models and at various evolutionary stages.}

\rev{\subsection{Total amount of SLR-rich material reaching the solar protoplanetary disk}\label{SN25}}

Here we discuss the total mass of materials, $M_{\rm tot}^{\rm SN\text{-}to\text{-}PPD}$, that is provided by the supernova ejecta and \rev{needs to be accreted onto the solar protoplanetary disk to explain the abundance of $^{26}$Al
observed  in the solar system. }

A supernova produces $^{26}$Al and the yield, $M_{\rm 26Al}^{\rm SN}$, is a function of the progenitor mass, $M_{\star}$.
\citet{PortegiesZwart2019} provided a simple fitting formula, 
\begin{equation}
    \log(M_{\rm 26Al}^{\rm SN} / \Msun) = 2.43 \log{( M_{\star} / \Msun )} - 7.23 \,. 
    \label{eq:PZ}
\end{equation}
The concentration of $^{26}$Al in the supernova ejecta, $C \MK{= M_{\rm 26Al}^{\rm SN} / M_{\rm ejecta}}$, is approximately given by $C \simeq M_{\rm 26Al}^{\rm SN} / M_{\star}$ because the mass of the remnant is negligibly smaller than that of the ejecta, $M_{\rm ejecta}$. 
Thus we can rewrite Eq.\,\ref{eq:PZ} as
\begin{equation}
    \log(C) \simeq \log(M_{\rm 26Al}^{\rm SN} / M_{\star}) = 1.43 \log{( M_{\star} / \Msun )} - 7.23 \,,
    \label{eq:PZ2}
\end{equation}
to estimate $C$ as a function of $M_{\star}$.
As the $^{26}$Al-poor CAIs are nearly free from $^{26}$Al (see Section\ \ref{sec.FUN}), we can assume that almost all $^{26}$Al in the protoplanetary disk is provided by an injection event from a nearby supernova.
We can, therefore, estimate the total amount of SN material reaching the PPD, $M_{\rm tot}^{\rm SN\text{-}to\text{-}PPD}$, as
\begin{equation}
M_{\rm tot}^{\rm SN\text{-}to\text{-}PPD} \simeq M_{\rm 26Al}^{\rm PPD} / C.
\label{eq:MSN}
\end{equation}
For a progenitor mass of $M_{\star} = 25\,\Msun$, 
Eqs.\,\ref{eq:PZ2} and \ref{eq:MSN} give $C = 5.9 \times 10^{-6}$ and $M_{\rm tot}^{\rm SN\text{-}to\text{-}PPD} \simeq 2 \times 10^{-6}\,\Msun$.
This corresponds to the expected amount of supernova ejecta accreted onto the solar protoplanetary disk with the required amount of $^{26}$Al to be compatible with the observations.

\rev{We note that $^{26}$Al is also produced by WR winds. According to Equation A.4 of \citet{PortegiesZwart2019}, however, the $^{26}$Al yield of a $25\,\Msun$ star is negligible compared to that of the SN. In this case, we can thus ignore  the contribution of WR winds  regarding the total production/ejection of $^{26}$Al for a $25\,\Msun$ star. }

The above derived  $M_{\rm tot}^{\rm SN\text{-}to\text{-}PPD}$ value,   
can also be used to get insights on the still debated $( M_{\rm 60Fe} / M_{\rm Fe} )_{\rm initial}$ isotopic ratio at the time of CAI formation
(cf. Section\,\ref{sec:SLR13}). %
For that we calculate the yield of $^{60}$Fe for a progenitor mass of 25\,\Msun\ using the relation provided by \citet{PortegiesZwart2019}, see their Eq.\,B.5. For a solar abundance of $^{56}$Fe of  ${( M_{\rm Fe} / M_{\rm H} )}_{\rm solar}\sim 3.2\times 10^{-5}$ \citep[e.g.,][]{Gounelle2015},
we found $( M_{\rm 60Fe} / M_{\rm Fe} )_{\rm initial}  \sim8\times 10^{-7}$. This value seems to be consistent with the old estimates suggesting a high abundance \citep[e.g.,][]{Mishra2016,Telus2018}, and inconsistent with the most recent lower abundance estimates \citep[e.g.,][]{Kodolanyi2022} as mentioned in Section\,\ref{sec:SLR13}.  
We however note multiple caveats in this estimated initial ($^{60}$Fe/$^{56}$Fe) isotope abundance ratio, which 1) assumes that all the $^{26}$Al is injected from a SN (and no contribution from WR winds) or there is no pre-existing $^{60}$Fe in the molecular core, 2) depends on the choice of the progenitor mass, and 3) the detailed SN explosion model.\\

\rev{As discussed in Section\,\ref{sec:SLR13}, the source of SLRs in the early solar system is still under debate. 
Accumulation of materials provided from WR winds may also be reasonable, especially, based on recent measurements of lower values of the initial abundances of $^{60}$Fe, which could be explained without the need of invoking a nearby SN explosion (cf. Section\,\ref{sec:SLR13} for references).}
\rev{For the WR wind enrichment scenario,  stars initially 
more massive than 40\,\Msun\ or 60\,\Msun\ may be needed. %
For example,
according to Equation A.4 of \citet{PortegiesZwart2019}, the WR winds of an $M_{\star}=60\,\Msun$ star may carry  
$M_{\rm 26Al}^{\rm WR}\sim2\times10^{-4}\,\Msun$ of $^{26}$Al.  
Such  high-mass stars lose a large fraction of their initial mass due to winds and may not explode in an SN. According to  stellar evolution models, About $90\%$ of the initial 60\,\Msun\ is lost in winds \citep[][]{Vuissoz2004,Sukhbold2016}.
Thus, the concentration of $^{26}$Al in the WR wind ejecta is $C^{\rm WR}=M_{\rm 26Al}^{\rm WR}/(0.9\,M_{\star})\sim4\times10^{-6}$. Consequently, the total amount of WR material reaching the PPD is $M_{\rm tot}^{\rm WR\text{-}to\text{-}PPD} \simeq M_{\rm 26Al}^{\rm PPD} / C^{\rm WR} \sim 2.5 \times 10^{-6}\,\Msun$. This is comparable to   $M_{\rm tot}^{\rm SN\text{-}to\text{-}PPD}$ for a $25\,\Msun$ progenitor.}\\

\rev{In the following section, we evaluate whether 
the necessary  ejecta mass 
$M_{\rm tot}^{\rm ejecta\text{-}to\text{-}PPD}\sim10^{-6}\,\Msun$
with SLR-rich material 
 reach  the PPD within the timescale of CAI formation},
and propose that the SLR-rich material is channeled by streamers within the filament onto the proto-Sun system.\\

\subsection{Mass accretion from the filament onto the proto-Sun system}

The \rev{SLR-rich} ejecta first encounters the filament, is mixed with the filament gas and then flows onto the solar protoplanetary disk.
Recent observations have shown the presence of streamers providing fresh material from the filament to the young stellar system (see Section\,\ref{sec:streamers} and Figure\,\ref{SummaryFig}).
These streamers may also channel the \rev{SLR-rich} ejecta onto the solar protoplanetary disk \rev{and contribute to increasing the cross section of the region that intercepts the ejecta. This cross section may thus not be limited solely to the prestellar core or PPD scales as assumed in earlier studies, but could be set by the filament properties, such as the filament length and width. 
The total mass of ejecta  intercepted by the filament is
}

\begin{figure}[!h]  
   \centering
  \resizebox{9.cm}{!}{ 
\includegraphics[angle=0]{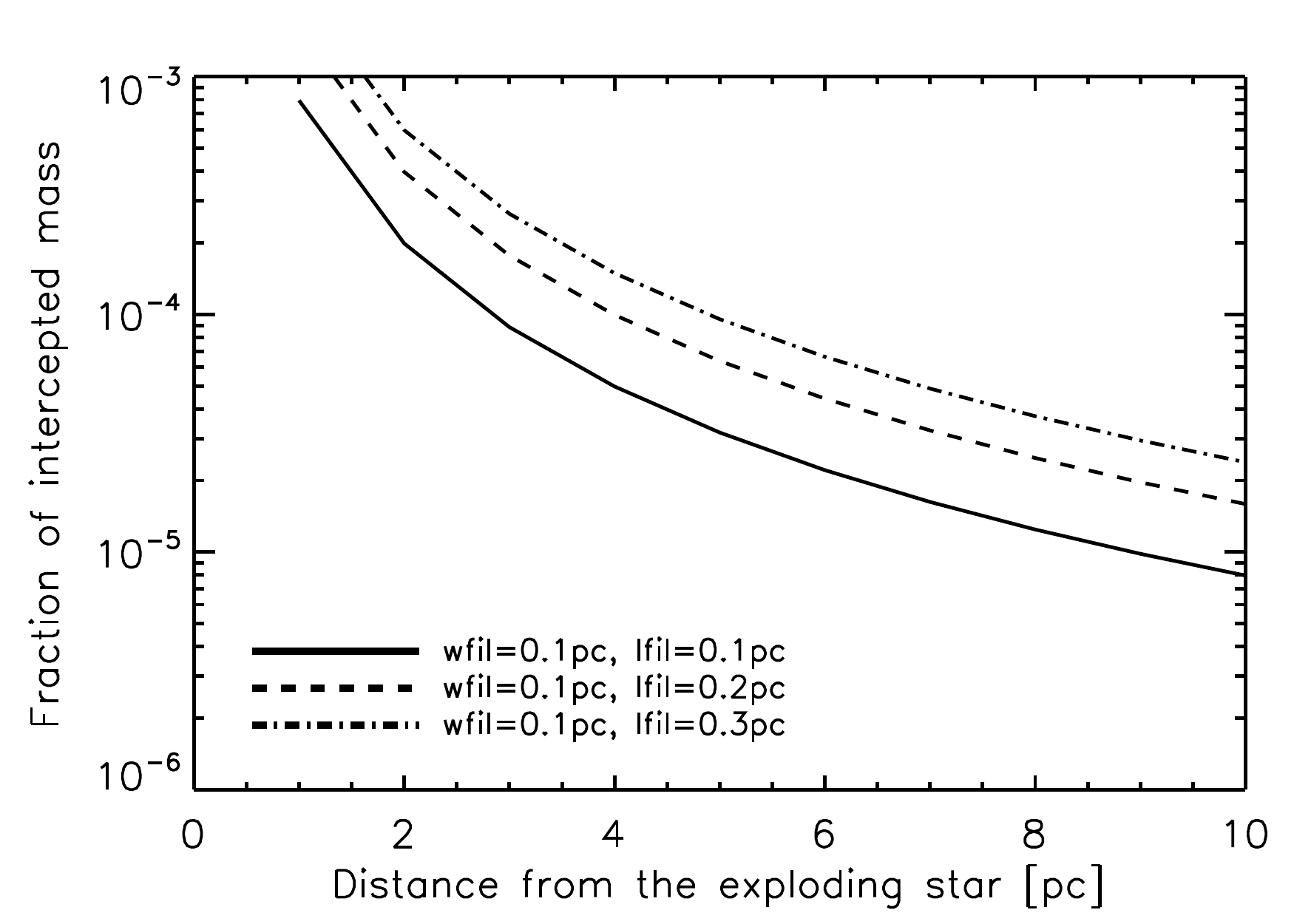}}
  \caption{\rev{Fraction of ejecta mass, $M_{\rm tot}^{\rm ejecta\text{-}to\text{-}fil}/M_{\rm ejecta}$,  intercepted by the filament against  the distance of the filament from the exploding star. The three curves correspond to three estimates of the filament cross section, $S_{\rm fil}=w_{\rm{fil}}\,l_{\rm{fil}}$  for $l_{\rm{fil}}=w_{\rm{fil}}, 2w_{\rm{fil}}$, and $3w_{\rm{fil}}$, where $l_{\rm{fil}}$ is the longitudinal extent and $w_{\rm{fil}}=W_{\rm{fil}}=0.1$\,pc the radial extent (cf. Section\,\ref{sec:HFS1}).  }}
  \label{Mfrac_d}
  \end{figure}

\rev{
\begin{equation}
M_{\rm tot}^{\rm ejecta\text{-}to\text{-}fil} = M_{\rm ejecta}\frac{S_{\rm{fil}}}{4\,\pi\,d^2},
\label{eq:ejectaFil}
\end{equation}
}
\rev{with $S_{\rm{fil}}$ the filament cross section
 and $d$ the distance %
 from the massive star.
 The filament cross section is $S_{\rm{fil}}=w_{\rm{fil}}\,l_{\rm{fil}}$, where $w_{\rm{fil}}$ and $l_{\rm{fil}}$ are the extent  across and along the cylindrical filament,  respectively.
Figure\,\ref{Mfrac_d} shows the parameter space of the distance of the exploding star vs. the fraction of ejecta mass, $M_{\rm tot}^{\rm ejecta\text{-}to\text{-}fil}/M_{\rm ejecta}$,  intercepted by the filament for three conservative estimates of the %
cross section. 
The ejecta mass intercepted by the filament $M_{\rm tot}^{\rm ejecta\text{-}to\text{-}fil}$, for   $M_{\rm ejecta}\sim25$--$60$\,\Msun, conservative cross sections, and $d\lesssim10\,$pc, is $>10^{-4}$\,\Msun, more than two orders of magnitude larger than the required captured mass by the PPD, $M_{\rm tot}^{\rm ejecta\text{-}to\text{-}PPD}\sim10^{-6}\,\Msun$ (cf. Section\,\ref{SN25}).} 

\rev{Various aspects can affect, however,  the  ejecta mass  
containing $^{26}$Al intercepted by the filament and transferred  to the PPD.}
\rev{ Considering  that $M_{\rm tot}^{\rm ejecta\text{-}to\text{-}fil}$ is }
channeled through  streamers onto the solar protoplanetary disk,  
the duration of the accretion along the streamers, $t_{\rm streamer}$, to provide the required amount of $^{26}$Al \rev{can be written as} 
\begin{equation}
t_{\rm streamer} = \frac{M_{\rm tot}^{\rm \rev{ejecta}\text{-}to\text{-}PPD}}{ w\,\dot{M}_{\rm streamer}},
\label{eq:tstr}
\end{equation}
\rev{%
where $\dot{M}_{\rm streamer} \sim 10^{-6}\,\Msun\,{\rm yr}^{-1}$ is the observed mass accretion rate along the streamers (see Section\,\ref{sec:streamers}).}
The $w$ factor in Eq.\,\ref{eq:tstr},  conceptually combines various aspects that affect the total mass accretion rate onto the Sun protostellar system, such as  1) the mass fraction of %
\rev{SLR-rich} ejecta in a given streamer, 2) the total number of streamers accreting onto the solar protoplanetary disk, 
and 3) the mass fraction of \rev{SLR-rich} ejecta injected onto the core and the disk, but not channeled by streamers.
As the duration of CAI formation is a few $10^{5}$ years or less (see Section\ \ref{sec.FUN}), injection of $^{26}$Al-rich materials through the streamer should be completed within the timescale of $0.1$ Myr.
Also, as estimated in Section\,\ref{SNe}, a Sun-like star forming filament may survive the SN blast for $>0.3$\,Myr, which would provide sufficient time for the SN ejecta to be channeled onto the solar protoplanetary disk and for the CAI to form in the disk within the timescale of $t_{\rm streamer} < 0.1$ Myrs. 
\rev{Hence, for a timescale of $0.1$\,Myr and $M_{\rm tot}^{\rm \rev{ejecta}\text{-}to\text{-}PPD}\sim 10^{-6}\,\Msun$ (cf. Section\,\ref{SN25}), $ w\sim10^{-5}$ is a plausible value to reproduce the observations. 
Assuming all the ejecta mass intercepted by the filament $M_{\rm tot}^{\rm \rev{ejecta}\text{-}to\text{-}PPD}=M_{\rm tot}^{\rm \rev{ejecta}\text{-}to\text{-}fil}\sim10^{-4}\,\Msun$ (for conservative cross sections and distances between the filament and the massive star) is transferred to the PPD in $0.1$\,Myr, 
$ w\sim10^{-3}$, which however results in an amount larger than  that required to explain the observations.}

\rev{Dedicated numerical simulations are needed to describe the interactions between the stellar feedback (WR winds and/or SNe) and the host filament, as well as the diffusion processes and timescale for the \rev{SLR-rich} ejecta to travel from  the filament, to  the streamers and onto the young solar system protostellar disk.}

\section{Summary and conclusions}\label{conclusion}

In this study, we discuss the importance of the hub-filament configuration where sun-like stars and massive stars form in the understanding of the observed abundances of SLRs, especially  $^{26}{\rm Al}$, in the CAIs of the solar system. 

As summarized in Fig.\,\ref{SummaryFig}, low- to intermediate-mass stars form along dense molecular filaments, which may merge to form a hub-filament system with a dense hub at the junction of multiple filaments (a). These  hubs provide the large mass and the high density %
required for the formation of high-mass stars. The feedback from these high-mass stars during their formation and main sequence phase drives   HII  bubbles that open holes in the dense hub and escape in the ISM sweeping up the dust and gas around the dense filaments connected to the hub (b). At an advance evolutionary stage the HII  regions erode the tip of the filaments close to the hub detaching the filaments from the hub. These dense filaments, such as a  $\sim90\,\sunpc$ filament hosting pre- and proto-Suns, are not destroyed but compressed by the expanding bubble. This compression enhances the density contrast between the filament and the ambient inter-filamentary medium  and enrich the filament with the ejecta from the winds from massive stars. 
At the end of its life, the most massive star formed in the hub, after a WR wind phase explodes in a supernova (c). 
A Sun-like star forming filament with a line mass of $\sim90\,\sunpc$   and density of $\sim10^5$\,cm$^{-3}$ (Section\,\ref{sec:HFS1}) may survive the SN shock for  $>0.3$\,Myr (Section\,\ref{SNe}) and be enriched by the ejecta from the SN containing SLRs such as  %
$^{26}$Al. The SN ejecta is channeled onto the young solar system enriching  the CAIs forming in the disk with SLRs within 0.1\,Myr %
 (Section\,\ref{Discussion}). These SLR-enriched gas and dust may be channeled along the filament from distances larger than the core size  and the protoplanetary disk size, as recent observations revealed (Section\,\ref{sec:streamers}). Such streamers have been observed at different protostellar  evolutionary stages %
 suggesting their possible role in providing material with varying amount of  SLRs during the entire evolution of the protostellar system and the formation of planetray systems (d).

Overall, we suggest that a proto-Sun forming in a supercritical filament next to a hub hosting massive stars can survive  both the early HII  region feedback from massive stars and 
the later 
violent feedback from a SN (preceded by a short WR phase), which will provide the required amount of $^{26}$Al to explain the observations. 
This scenario may have multiple 
important implications 
in our understanding of the formation, evolution, and properties of stellar systems. For example, the host filament may  play an important role in shielding the young solar system from the far-ultraviolet radiation from OB stars that would  photoevaporate the protostellar disk affecting its final size, which would  have a direct impact on planet formation within the disk \citep[e.g.,][]{Adams2004}. 
Mass segregated star cluster formation in hub-filament configurations may also affect the dynamical evolution of the stellar cluster and its impact on the truncation of the disk size of the stellar systems \citep[e.g.,][]{Kobayashi2001,Bhandare2016} formed along the filaments.
We thus conclude that considering hub-filament configurations as the birth environment of the Sun is important when discussing interpretations of the observed properties of the solar system, e.g., isotope ratio, disk size, number of stars of the host cluster. 

To assess the reliability of our scenario, dedicated magnetohydrodynamical numerical simulations are required to quantitatively constrain the impact  of  HII  region, \rev{WR winds,}  and SN feedback on the surrounding filaments and the distruction timescales of the star-forming filaments. Numerical simulations are also desirable to estimate diffusion timescales of the SLR-rich material from the filaments to the streamers to the protostellar disk. 
We also anticipate future observations to better describe the properties of the streamers as well as theoretical studies to understand the origin of these streamers and their impact on the mass accretion on young stellar systems.

\begin{acknowledgements}
We thank Pedro Palmeirim for fruitful discussions at the early phase of this study.
We also thank Tomoya Takiwaki \DArr{and Jeong-Gyu Kim} for their comments.
S.A.\ was supported by JSPS KAKENHI Grant No.\ JP20J00598.
S.M.\ was supported by JSPS KAKENHI Grant No.\ JP21J00086 and JP22K14081.
Y.H.\ was supported by JSPS KAKENHI Grants No.\ JP21J00153, JP20K14532, JP21H04499, JP21K03614, and JP22H01259.
E.K.\ was supported by JSPS KAKENHI Grant No.\ 18H05438.
  \end{acknowledgements}






\bibliographystyle{aasjournal}
\bibliography{bibfile}

\begin{thebibliography}{}
\expandafter\ifx\csname natexlab\endcsname\relax\def\natexlab#1{#1}\fi
\providecommand{\url}[1]{\href{#1}{#1}}
\providecommand{\dodoi}[1]{doi:~\href{http://doi.org/#1}{\nolinkurl{#1}}}
\providecommand{\doeprint}[1]{\href{http://ascl.net/#1}{\nolinkurl{http://ascl.net/#1}}}
\providecommand{\doarXiv}[1]{\href{https://arxiv.org/abs/#1}{\nolinkurl{https://arxiv.org/abs/#1}}}

\bibitem[{{Adams}(2010)}]{Adams2010}
{Adams}, F.~C. 2010, \araa, 48, 47, \dodoi{10.1146/annurev-astro-081309-130830}

\bibitem[{{Adams} {et~al.}(2004){Adams}, {Hollenbach}, {Laughlin}, \&
  {Gorti}}]{Adams2004}
{Adams}, F.~C., {Hollenbach}, D., {Laughlin}, G., \& {Gorti}, U. 2004, \apj,
  611, 360, \dodoi{10.1086/421989}

\bibitem[{{Alves} {et~al.}(2001){Alves}, {Lada}, \& {Lada}}]{Alves2001}
{Alves}, J.~F., {Lada}, C.~J., \& {Lada}, E.~A. 2001, Nature, 409, 159

\bibitem[{{Amelin} {et~al.}(2002){Amelin}, {Krot}, {Hutcheon}, \&
  {Ulyanov}}]{Amelin2002}
{Amelin}, Y., {Krot}, A.~N., {Hutcheon}, I.~D., \& {Ulyanov}, A.~A. 2002,
  Science, 297, 1678, \dodoi{10.1126/science.1073950}

\bibitem[{{Andr{\'e}} {et~al.}(2019){Andr{\'e}}, {Arzoumanian}, {K{\"o}nyves},
  {Shimajiri}, \& {Palmeirim}}]{Andre2019}
{Andr{\'e}}, P., {Arzoumanian}, D., {K{\"o}nyves}, V., {Shimajiri}, Y., \&
  {Palmeirim}, P. 2019, {A{\&}A}, 629, L4, \dodoi{10.1051/0004-6361/201935915}

\bibitem[{{Andr{\'e}} {et~al.}(2014){Andr{\'e}}, {Di Francesco},
  {Ward-Thompson}, {Inutsuka}, {Pudritz}, \& {Pineda}}]{Andre2014}
{Andr{\'e}}, P., {Di Francesco}, J., {Ward-Thompson}, D., {et~al.} 2014,
  Protostars and Planets VI, 27,
  \dodoi{10.2458/azu_uapress_9780816531240-ch002}

\bibitem[{{Andre} {et~al.}(2000){Andre}, {Ward-Thompson}, \&
  {Barsony}}]{Andre2000}
{Andre}, P., {Ward-Thompson}, D., \& {Barsony}, M. 2000, Protostars and Planets
  IV, 59

\bibitem[{{Andr{\'e}} {et~al.}(2010){Andr{\'e}}, {Men'shchikov}, {Bontemps},
  {K{\"o}nyves}, {Motte}, {Schneider}, {Didelon}, {Minier}, {Saraceno},
  {Ward-Thompson}, {di Francesco}, {White}, {Molinari}, {Testi}, {Abergel},
  {Griffin}, {Henning}, {Royer}, {Mer{\'{\i}}n}, {Vavrek}, {Attard},
  {Arzoumanian}, {Wilson}, {Ade}, {Aussel}, {Baluteau}, {Benedettini},
  {Bernard}, {Blommaert}, {Cambr{\'e}sy}, {Cox}, {di Giorgio}, {Hargrave},
  {Hennemann}, {Huang}, {Kirk}, {Krause}, {Launhardt}, {Leeks}, {Le Pennec},
  {Li}, {Martin}, {Maury}, {Olofsson}, {Omont}, {Peretto}, {Pezzuto}, {Prusti},
  {Roussel}, {Russeil}, {Sauvage}, {Sibthorpe}, {Sicilia-Aguilar}, {Spinoglio},
  {Waelkens}, {Woodcraft}, \& {Zavagno}}]{Andre2010}
{Andr{\'e}}, P., {Men'shchikov}, A., {Bontemps}, S., {et~al.} 2010, A{\&}A,
  518, L102, \dodoi{10.1051/0004-6361/201014666}

\bibitem[{{Andr{\'e}} {et~al.}(2022){Andr{\'e}}, {Palmeirim}, \&
  {Arzoumanian}}]{Andre2022}
{Andr{\'e}}, P.~J., {Palmeirim}, P., \& {Arzoumanian}, D. 2022, \aap, 667, L1,
  \dodoi{10.1051/0004-6361/202244541}

\bibitem[{{Arakawa} \& {Kokubo}(2022)}]{Arakawa2022}
{Arakawa}, S., \& {Kokubo}, E. 2022, arXiv e-prints, arXiv:2212.13772,
  \dodoi{10.48550/arXiv.2212.13772}

\bibitem[{{Arthur} {et~al.}(2011){Arthur}, {Henney}, {Mellema}, {de Colle}, \&
  {V{\'a}zquez-Semadeni}}]{Arthur2011}
{Arthur}, S.~J., {Henney}, W.~J., {Mellema}, G., {de Colle}, F., \&
  {V{\'a}zquez-Semadeni}, E. 2011, \mnras, 414, 1747,
  \dodoi{10.1111/j.1365-2966.2011.18507.x}

\bibitem[{{Arzoumanian} {et~al.}(2013){Arzoumanian}, {Andr{\'e}}, {Peretto}, \&
  {K{\"o}nyves}}]{Arzoumanian2013}
{Arzoumanian}, D., {Andr{\'e}}, P., {Peretto}, N., \& {K{\"o}nyves}, V. 2013,
  {A{\&}A}, 553, A119, \dodoi{10.1051/0004-6361/201220822}

\bibitem[{{Arzoumanian} {et~al.}(2011){Arzoumanian}, {Andr{\'e}}, {Didelon},
  {K{\"o}nyves}, {Schneider}, {Men'shchikov}, {Sousbie}, {Zavagno}, {Bontemps},
  {di Francesco}, {Griffin}, {Hennemann}, {Hill}, {Kirk}, {Martin}, {Minier},
  {Molinari}, {Motte}, {Peretto}, {Pezzuto}, {Spinoglio}, {Ward-Thompson},
  {White}, \& {Wilson}}]{Arzoumanian2011}
{Arzoumanian}, D., {Andr{\'e}}, P., {Didelon}, P., {et~al.} 2011, A{\&}A, 529,
  L6, \dodoi{10.1051/0004-6361/201116596}

\bibitem[{{Arzoumanian} {et~al.}(2019){Arzoumanian}, {Andr{\'e}},
  {K{\"o}nyves}, {Palmeirim}, {Roy}, {Schneider}, {Benedettini}, {Didelon}, {Di
  Francesco}, {Kirk}, \& {Ladjelate}}]{Arzoumanian2019}
{Arzoumanian}, D., {Andr{\'e}}, P., {K{\"o}nyves}, V., {et~al.} 2019, A{\&}A,
  621, A42.
\newblock \doarXiv{1810.00721}

\bibitem[{{Asplund} {et~al.}(2021){Asplund}, {Amarsi}, \&
  {Grevesse}}]{Asplund2021}
{Asplund}, M., {Amarsi}, A.~M., \& {Grevesse}, N. 2021, \aap, 653, A141,
  \dodoi{10.1051/0004-6361/202140445}

\bibitem[{{Bergin} \& {Tafalla}(2007)}]{Bergin2007}
{Bergin}, E.~A., \& {Tafalla}, M. 2007, AR{\&}A, 45, 339,
  \dodoi{10.1146/annurev.astro.45.071206.100404}

\bibitem[{{Bhandare} {et~al.}(2016){Bhandare}, {Breslau}, \&
  {Pfalzner}}]{Bhandare2016}
{Bhandare}, A., {Breslau}, A., \& {Pfalzner}, S. 2016, \aap, 594, A53,
  \dodoi{10.1051/0004-6361/201628086}

\bibitem[{{Bonnor}(1956)}]{Bonnor1956}
{Bonnor}, W.~B. 1956, MNRAS, 116, 351

\bibitem[{{Boss}(2012)}]{Boss2012}
{Boss}, A.~P. 2012, Annual Review of Earth and Planetary Sciences, 40, 23,
  \dodoi{10.1146/annurev-earth-042711-105552}

\bibitem[{{Cameron} \& {Truran}(1977)}]{Cameron1977}
{Cameron}, A.~G.~W., \& {Truran}, J.~W. 1977, \icarus, 30, 447,
  \dodoi{10.1016/0019-1035(77)90101-4}

\bibitem[{{Caselli} {et~al.}(2002){Caselli}, {Benson}, {Myers}, \&
  {Tafalla}}]{Caselli2002}
{Caselli}, P., {Benson}, P.~J., {Myers}, P.~C., \& {Tafalla}, M. 2002, ApJ,
  572, 238, \dodoi{10.1086/340195}

\bibitem[{{Chen} {et~al.}(2019){Chen}, {Storm}, {Li}, {Mundy}, {Frayer}, {Li},
  {Church}, {Friesen}, {Harris}, {Looney}, {Offner}, {Ostriker}, {Pineda},
  {Tobin}, \& {Chen}}]{Chen2019}
{Chen}, C.-Y., {Storm}, S., {Li}, Z.-Y., {et~al.} 2019, \mnras, 490, 527,
  \dodoi{10.1093/mnras/stz2633}

\bibitem[{{Clayton}(1977)}]{Clayton1977}
{Clayton}, D.~D. 1977, \icarus, 32, 255, \dodoi{10.1016/0019-1035(77)90001-X}

\bibitem[{{Close} \& {Pittard}(2017)}]{Close2017}
{Close}, J.~L., \& {Pittard}, J.~M. 2017, \mnras, 469, 1117,
  \dodoi{10.1093/mnras/stx897}

\bibitem[{{Connelly} {et~al.}(2012){Connelly}, {Bizzarro}, {Krot}, {Nordlund},
  {Wielandt}, \& {Ivanova}}]{Connelly2012}
{Connelly}, J.~N., {Bizzarro}, M., {Krot}, A.~N., {et~al.} 2012, Science, 338,
  651, \dodoi{10.1126/science.1226919}

\bibitem[{{Dale} {et~al.}(2013){Dale}, {Ngoumou}, {Ercolano}, \&
  {Bonnell}}]{Dale2013}
{Dale}, J.~E., {Ngoumou}, J., {Ercolano}, B., \& {Bonnell}, I.~A. 2013, \mnras,
  436, 3430, \dodoi{10.1093/mnras/stt1822}

\bibitem[{{Dauphas} \& {Chaussidon}(2011)}]{Dauphas2011}
{Dauphas}, N., \& {Chaussidon}, M. 2011, Annual Review of Earth and Planetary
  Sciences, 39, 351, \dodoi{10.1146/annurev-earth-040610-133428}

\bibitem[{{Deharveng} {et~al.}(2015){Deharveng}, {Zavagno}, {Samal},
  {Anderson}, {LeLeu}, {Brevot}, {Duarte-Cabral}, {Molinari}, {Pestalozzi},
  {Foster}, {Rathborne}, \& {Jackson}}]{Deharveng2015}
{Deharveng}, L., {Zavagno}, A., {Samal}, M.~R., {et~al.} 2015, \aap, 582, A1,
  \dodoi{10.1051/0004-6361/201423835}

\bibitem[{{Desch} {et~al.}(2004){Desch}, {Connolly}, \&
  {Srinivasan}}]{Desch2004}
{Desch}, S.~J., {Connolly}, Harold~C., J., \& {Srinivasan}, G. 2004, \apj, 602,
  528, \dodoi{10.1086/380831}

\bibitem[{{Dunham} {et~al.}(2022){Dunham}, {Wadhwa}, {Desch}, {Liu}, {Fukuda},
  {Kita}, {Hertwig}, {Hervig}, {Defouilloy}, {Simon}, {Davidson}, {Schrader},
  \& {Fujimoto}}]{Dunham2022}
{Dunham}, E.~T., {Wadhwa}, M., {Desch}, S.~J., {et~al.} 2022, \gca, 324, 194,
  \dodoi{10.1016/j.gca.2022.02.002}

\bibitem[{{Dwarkadas} {et~al.}(2017){Dwarkadas}, {Dauphas}, {Meyer},
  {Boyajian}, \& {Bojazi}}]{Dwarkadas2017}
{Dwarkadas}, V.~V., {Dauphas}, N., {Meyer}, B., {Boyajian}, P., \& {Bojazi}, M.
  2017, \apj, 851, 147, \dodoi{10.3847/1538-4357/aa992e}

\bibitem[{{Fujimoto} {et~al.}(2018){Fujimoto}, {Krumholz}, \&
  {Tachibana}}]{Fujimoto2018}
{Fujimoto}, Y., {Krumholz}, M.~R., \& {Tachibana}, S. 2018, \mnras, 480, 4025,
  \dodoi{10.1093/mnras/sty2132}

\bibitem[{{Fukai} \& {Arakawa}(2021)}]{Fukai2021}
{Fukai}, R., \& {Arakawa}, S. 2021, \apj, 908, 64

\bibitem[{{Fukuda} {et~al.}(2021){Fukuda}, {Hiyagon}, {Fujiya}, {Kagoshima},
  {Itano}, {Iizuka}, {Kita}, \& {Sano}}]{Fukuda2021}
{Fukuda}, K., {Hiyagon}, H., {Fujiya}, W., {et~al.} 2021, \gca, 293, 187,
  \dodoi{10.1016/j.gca.2020.10.011}

\bibitem[{{Gaches} {et~al.}(2020){Gaches}, {Walch}, {Offner}, \&
  {M{\"u}nker}}]{Gaches2020}
{Gaches}, B. A.~L., {Walch}, S., {Offner}, S. S.~R., \& {M{\"u}nker}, C. 2020,
  \apj, 898, 79, \dodoi{10.3847/1538-4357/ab9a38}

\bibitem[{{Georgy} {et~al.}(2012){Georgy}, {Ekstr{\"o}m}, {Meynet}, {Massey},
  {Levesque}, {Hirschi}, {Eggenberger}, \& {Maeder}}]{Georgy2012}
{Georgy}, C., {Ekstr{\"o}m}, S., {Meynet}, G., {et~al.} 2012, \aap, 542, A29,
  \dodoi{10.1051/0004-6361/201118340}

\bibitem[{{Gounelle}(2015)}]{Gounelle2015}
{Gounelle}, M. 2015, \aap, 582, A26, \dodoi{10.1051/0004-6361/201526174}

\bibitem[{{Gounelle} \& {Meynet}(2012)}]{Gounelle2012}
{Gounelle}, M., \& {Meynet}, G. 2012, \aap, 545, A4,
  \dodoi{10.1051/0004-6361/201219031}

\bibitem[{{Gritschneder} {et~al.}(2012){Gritschneder}, {Lin}, {Murray}, {Yin},
  \& {Gong}}]{Gritschneder2012}
{Gritschneder}, M., {Lin}, D.~N.~C., {Murray}, S.~D., {Yin}, Q.~Z., \& {Gong},
  M.~N. 2012, \apj, 745, 22, \dodoi{10.1088/0004-637X/745/1/22}

\bibitem[{{Hanawa} {et~al.}(2022){Hanawa}, {Sakai}, \& {Yamamoto}}]{Hanawa2022}
{Hanawa}, T., {Sakai}, N., \& {Yamamoto}, S. 2022, \apj, 932, 122,
  \dodoi{10.3847/1538-4357/ac6e6a}

\bibitem[{{Hartmann} {et~al.}(2016){Hartmann}, {Herczeg}, \&
  {Calvet}}]{Hartmann2016}
{Hartmann}, L., {Herczeg}, G., \& {Calvet}, N. 2016, \araa, 54, 135,
  \dodoi{10.1146/annurev-astro-081915-023347}

\bibitem[{{Holst} {et~al.}(2013){Holst}, {Olsen}, {Paton}, {Nagashima},
  {Schiller}, {Wielandt}, {Larsen}, {Connelly}, {J{\o}rgensen}, {Krot},
  {Nordlund}, \& {Bizzarro}}]{Holst2013}
{Holst}, J.~C., {Olsen}, M.~B., {Paton}, C., {et~al.} 2013, Proceedings of the
  National Academy of Science, 110, 8819, \dodoi{10.1073/pnas.1300383110}

\bibitem[{{Hosokawa} \& {Inutsuka}(2006)}]{Hosokawa2006}
{Hosokawa}, T., \& {Inutsuka}, S.-i. 2006, \apj, 646, 240,
  \dodoi{10.1086/504789}

\bibitem[{{Huss} {et~al.}(2009){Huss}, {Meyer}, {Srinivasan}, {Goswami}, \&
  {Sahijpal}}]{Huss2009}
{Huss}, G.~R., {Meyer}, B.~S., {Srinivasan}, G., {Goswami}, J.~N., \&
  {Sahijpal}, S. 2009, \gca, 73, 4922, \dodoi{10.1016/j.gca.2009.01.039}

\bibitem[{{Inutsuka} \& {Miyama}(1997)}]{Inutsuka1997}
{Inutsuka}, S., \& {Miyama}, S.~M. 1997, ApJ, 480, 681, \dodoi{10.1086/303982}

\bibitem[{{Inutsuka} {et~al.}(2015){Inutsuka}, {Inoue}, {Iwasaki}, \&
  {Hosokawa}}]{Inutsuka2015}
{Inutsuka}, S.-i., {Inoue}, T., {Iwasaki}, K., \& {Hosokawa}, T. 2015,
  {A{\&}A}, 580, A49, \dodoi{10.1051/0004-6361/201425584}

\bibitem[{{Jacquet}(2019)}]{Jacquet2019}
{Jacquet}, E. 2019, \aap, 624, A131, \dodoi{10.1051/0004-6361/201834754}

\bibitem[{{Jones} {et~al.}(2019){Jones}, {M{\"o}ller}, {Fryer}, {Fontes},
  {Trappitsch}, {Even}, {Couture}, {Mumpower}, \& {Safi-Harb}}]{Jones2019}
{Jones}, S.~W., {M{\"o}ller}, H., {Fryer}, C.~L., {et~al.} 2019, \mnras, 485,
  4287, \dodoi{10.1093/mnras/stz536}

\bibitem[{{Kennicutt}(1984)}]{Kennicutt1984}
{Kennicutt}, R.~C., J. 1984, \apj, 287, 116, \dodoi{10.1086/162669}

\bibitem[{{Kim} \& {Ostriker}(2015)}]{Kim2015}
{Kim}, C.-G., \& {Ostriker}, E.~C. 2015, \apj, 802, 99,
  \dodoi{10.1088/0004-637X/802/2/99}

\bibitem[{{Kim} {et~al.}(2018){Kim}, {Kim}, \& {Ostriker}}]{JGKim2018}
{Kim}, J.-G., {Kim}, W.-T., \& {Ostriker}, E.~C. 2018, \apj, 859, 68,
  \dodoi{10.3847/1538-4357/aabe27}

\bibitem[{{Kinoshita} {et~al.}(2021){Kinoshita}, {Nakamura}, \&
  {Wu}}]{Kinoshita2021}
{Kinoshita}, S.~W., {Nakamura}, F., \& {Wu}, B. 2021, \apj, 921, 150,
  \dodoi{10.3847/1538-4357/ac1d4b}

\bibitem[{{Klein} {et~al.}(1994){Klein}, {McKee}, \& {Colella}}]{Klein1994}
{Klein}, R.~I., {McKee}, C.~F., \& {Colella}, P. 1994, \apj, 420, 213,
  \dodoi{10.1086/173554}

\bibitem[{{Kobayashi} \& {Ida}(2001)}]{Kobayashi2001}
{Kobayashi}, H., \& {Ida}, S. 2001, \icarus, 153, 416,
  \dodoi{10.1006/icar.2001.6700}

\bibitem[{{Kodol{\'a}nyi} {et~al.}(2022{\natexlab{a}}){Kodol{\'a}nyi}, {Hoppe},
  {Vollmer}, {Berndt}, \& {M{\"u}ller}}]{Kodolanyi2022}
{Kodol{\'a}nyi}, J., {Hoppe}, P., {Vollmer}, C., {Berndt}, J., \& {M{\"u}ller},
  M. 2022{\natexlab{a}}, \apj, 929, 107, \dodoi{10.3847/1538-4357/ac5910}

\bibitem[{{Kodol{\'a}nyi} {et~al.}(2022{\natexlab{b}}){Kodol{\'a}nyi}, {Hoppe},
  {Vollmer}, {Berndt}, \& {M{\"u}ller}}]{Kodolanyi2022New}
---. 2022{\natexlab{b}}, \apj, 940, 95, \dodoi{10.3847/1538-4357/ac8b85}

\bibitem[{{K{\"o}nyves} {et~al.}(2015){K{\"o}nyves}, {Andr{\'e}},
  {Men'shchikov}, {Palmeirim}, {Arzoumanian}, {Schneider}, {Roy}, {Didelon},
  {Maury}, {Shimajiri}, {Di Francesco}, {Bontemps}, {Peretto}, {Benedettini},
  {Bernard}, {Elia}, {Griffin}, {Hill}, {Kirk}, {Ladjelate}, {Marsh}, {Martin},
  {Motte}, {Nguy{\^e}n Luong}, {Pezzuto}, {Roussel}, {Rygl}, {Sadavoy},
  {Schisano}, {Spinoglio}, {Ward-Thompson}, \& {White}}]{Konyves2015}
{K{\"o}nyves}, V., {Andr{\'e}}, P., {Men'shchikov}, A., {et~al.} 2015,
  {A{\&}A}, 584, A91, \dodoi{10.1051/0004-6361/201525861}

\bibitem[{{K{\"o}nyves} {et~al.}(2020){K{\"o}nyves}, {Andr{\'e}},
  {Arzoumanian}, {Schneider}, {Men'shchikov}, {Bontemps}, {Ladjelate},
  {Didelon}, {Pezzuto}, {Benedettini}, {Bracco}, {Di Francesco}, {Goodwin},
  {Rygl}, {Shimajiri}, {Spinoglio}, {Ward-Thompson}, \& {White}}]{Konyves2020}
{K{\"o}nyves}, V., {Andr{\'e}}, P., {Arzoumanian}, D., {et~al.} 2020, \aap,
  635, A34, \dodoi{10.1051/0004-6361/201834753}

\bibitem[{{K{\"o}{\"o}p} {et~al.}(2016){K{\"o}{\"o}p}, {Davis}, {Nakashima},
  {Park}, {Krot}, {Nagashima}, {Tenner}, {Heck}, \& {Kita}}]{Koeoep2016}
{K{\"o}{\"o}p}, L., {Davis}, A.~M., {Nakashima}, D., {et~al.} 2016, \gca, 189,
  70, \dodoi{10.1016/j.gca.2016.05.014}

\bibitem[{Krot(2019)}]{Krot2019}
Krot, A.~N. 2019, Meteoritics \& Planetary Science, 54, 1647,
  \dodoi{https://doi.org/10.1111/maps.13350}

\bibitem[{{Krumholz} {et~al.}(2014){Krumholz}, {Bate}, {Arce}, {Dale},
  {Gutermuth}, {Klein}, {Li}, {Nakamura}, \& {Zhang}}]{Krumholz2014}
{Krumholz}, M.~R., {Bate}, M.~R., {Arce}, H.~G., {et~al.} 2014, in Protostars
  and Planets VI, ed. H.~{Beuther}, R.~S. {Klessen}, C.~P. {Dullemond}, \&
  T.~{Henning}, 243--266, \dodoi{10.2458/azu_uapress_9780816531240-ch011}

\bibitem[{{Kuffmeier} {et~al.}(2021){Kuffmeier}, {Dullemond}, {Reissl}, \&
  {Goicovic}}]{Kuffmeier2021}
{Kuffmeier}, M., {Dullemond}, C.~P., {Reissl}, S., \& {Goicovic}, F.~G. 2021,
  \aap, 656, A161, \dodoi{10.1051/0004-6361/202039614}

\bibitem[{{Kuffmeier} {et~al.}(2020){Kuffmeier}, {Goicovic}, \&
  {Dullemond}}]{Kuffmeier2020}
{Kuffmeier}, M., {Goicovic}, F.~G., \& {Dullemond}, C.~P. 2020, \aap, 633, A3,
  \dodoi{10.1051/0004-6361/201936820}

\bibitem[{{Kumar} {et~al.}(2022){Kumar}, {Arzoumanian}, {Men'shchikov},
  {Palmeirim}, {Matsumura}, \& {Inutsuka}}]{Kumar2022}
{Kumar}, M.~S.~N., {Arzoumanian}, D., {Men'shchikov}, A., {et~al.} 2022, \aap,
  658, A114, \dodoi{10.1051/0004-6361/202140363}

\bibitem[{{Kumar} {et~al.}(2020){Kumar}, {Palmeirim}, {Arzoumanian}, \&
  {Inutsuka}}]{Kumar2020}
{Kumar}, M.~S.~N., {Palmeirim}, P., {Arzoumanian}, D., \& {Inutsuka}, S.~I.
  2020, \aap, 642, A87, \dodoi{10.1051/0004-6361/202038232}

\bibitem[{{Lee} {et~al.}(1998){Lee}, {Shu}, {Shang}, {Glassgold}, \&
  {Rehm}}]{Lee1998}
{Lee}, T., {Shu}, F.~H., {Shang}, H., {Glassgold}, A.~E., \& {Rehm}, K.~E.
  1998, \apj, 506, 898

\bibitem[{{Lugaro} {et~al.}(2018){Lugaro}, {Ott}, \& {Kereszturi}}]{Lugaro2018}
{Lugaro}, M., {Ott}, U., \& {Kereszturi}, {\'A}. 2018, Progress in Particle and
  Nuclear Physics, 102, 1, \dodoi{10.1016/j.ppnp.2018.05.002}

\bibitem[{{MacPherson} {et~al.}(2012){MacPherson}, {Kita}, {Ushikubo},
  {Bullock}, \& {Davis}}]{MacPherson2012}
{MacPherson}, G.~J., {Kita}, N.~T., {Ushikubo}, T., {Bullock}, E.~S., \&
  {Davis}, A.~M. 2012, Earth and Planetary Science Letters, 331, 43,
  \dodoi{10.1016/j.epsl.2012.03.010}

\bibitem[{{Mishra} {et~al.}(2016){Mishra}, {Marhas}, \& {Sameer}}]{Mishra2016}
{Mishra}, R.~K., {Marhas}, K.~K., \& {Sameer}. 2016, Earth and Planetary
  Science Letters, 436, 71, \dodoi{10.1016/j.epsl.2015.12.007}

\bibitem[{{Molinari} {et~al.}(2010){Molinari}, {Swinyard}, {Bally}, {Barlow},
  {Bernard}, {Martin}, {Moore}, {Noriega-Crespo}, {Plume}, {Testi}, {Zavagno},
  {Abergel}, {Ali}, {Anderson}, {Andr{\'e}}, {Baluteau}, {Battersby},
  {Beltr{\'a}n}, {Benedettini}, {Billot}, {Blommaert}, {Bontemps}, {Boulanger},
  {Brand}, {Brunt}, {Burton}, {Calzoletti}, {Carey}, {Caselli}, {Cesaroni},
  {Cernicharo}, {Chakrabarti}, {Chrysostomou}, {Cohen}, {Compiegne}, {de
  Bernardis}, {de Gasperis}, {di Giorgio}, {Elia}, {Faustini}, {Flagey},
  {Fukui}, {Fuller}, {Ganga}, {Garcia-Lario}, {Glenn}, {Goldsmith}, {Griffin},
  {Hoare}, {Huang}, {Ikhenaode}, {Joblin}, {Joncas}, {Juvela}, {Kirk},
  {Lagache}, {Li}, {Lim}, {Lord}, {Marengo}, {Marshall}, {Masi}, {Massi},
  {Matsuura}, {Minier}, {Miville-Desch{\^e}nes}, {Montier}, {Morgan}, {Motte},
  {Mottram}, {M{\"u}ller}, {Natoli}, {Neves}, {Olmi}, {Paladini}, {Paradis},
  {Parsons}, {Peretto}, {Pestalozzi}, {Pezzuto}, {Piacentini}, {Piazzo},
  {Polychroni}, {Pomar{\`e}s}, {Popescu}, {Reach}, {Ristorcelli}, {Robitaille},
  {Robitaille}, {Rod{\'o}n}, {Roy}, {Royer}, {Russeil}, {Saraceno}, {Sauvage},
  {Schilke}, {Schisano}, {Schneider}, {Schuller}, {Schulz}, {Sibthorpe},
  {Smith}, {Smith}, {Spinoglio}, {Stamatellos}, {Strafella}, {Stringfellow},
  {Sturm}, {Taylor}, {Thompson}, {Traficante}, {Tuffs}, {Umana}, {Valenziano},
  {Vavrek}, {Veneziani}, {Viti}, {Waelkens}, {Ward-Thompson}, {White},
  {Wilcock}, {Wyrowski}, {Yorke}, \& {Zhang}}]{Molinari2010}
{Molinari}, S., {Swinyard}, B., {Bally}, J., {et~al.} 2010, A{\&}A, 518, L100,
  \dodoi{10.1051/0004-6361/201014659}

\bibitem[{{Myers}(2009)}]{Myers2009}
{Myers}, P.~C. 2009, ApJ, 700, 1609, \dodoi{10.1088/0004-637X/700/2/1609}

\bibitem[{{Nakamura} {et~al.}(2006){Nakamura}, {McKee}, {Klein}, \&
  {Fisher}}]{Nakamura2006}
{Nakamura}, F., {McKee}, C.~F., {Klein}, R.~I., \& {Fisher}, R.~T. 2006, \apjs,
  164, 477, \dodoi{10.1086/501530}

\bibitem[{{Nittler} \& {Ciesla}(2016)}]{Nittler2016}
{Nittler}, L.~R., \& {Ciesla}, F. 2016, \araa, 54, 53,
  \dodoi{10.1146/annurev-astro-082214-122505}

\bibitem[{{Ouellette} {et~al.}(2010){Ouellette}, {Desch}, \&
  {Hester}}]{Ouellette2010}
{Ouellette}, N., {Desch}, S.~J., \& {Hester}, J.~J. 2010, \apj, 711, 597,
  \dodoi{10.1088/0004-637X/711/2/597}

\bibitem[{{Palmeirim} {et~al.}(2013){Palmeirim}, {Andr{\'e}}, {Kirk},
  {Ward-Thompson}, {Arzoumanian}, {K{\"o}nyves}, {Didelon}, {Schneider},
  {Benedettini}, {Bontemps}, {Di Francesco}, {Elia}, {Griffin}, {Hennemann},
  {Hill}, {Martin}, {Men'shchikov}, {Molinari}, {Motte}, {Nguyen Luong},
  {Nutter}, {Peretto}, {Pezzuto}, {Roy}, {Rygl}, {Spinoglio}, \&
  {White}}]{Palmeirim2013}
{Palmeirim}, P., {Andr{\'e}}, P., {Kirk}, J., {et~al.} 2013, {A{\&}A}, 550,
  A38, \dodoi{10.1051/0004-6361/201220500}

\bibitem[{{Park} {et~al.}(2017){Park}, {Nagashima}, {Krot}, {Huss}, {Davis}, \&
  {Bizzarro}}]{Park2017}
{Park}, C., {Nagashima}, K., {Krot}, A.~N., {et~al.} 2017, \gca, 201, 6,
  \dodoi{10.1016/j.gca.2016.10.002}

\bibitem[{{Peretto} {et~al.}(2013){Peretto}, {Fuller}, {Duarte-Cabral},
  {Avison}, {Hennebelle}, {Pineda}, {Andr{\'e}}, {Bontemps}, {Motte},
  {Schneider}, \& {Molinari}}]{Peretto2013}
{Peretto}, N., {Fuller}, G.~A., {Duarte-Cabral}, A., {et~al.} 2013, \aap, 555,
  A112, \dodoi{10.1051/0004-6361/201321318}

\bibitem[{{Peretto} {et~al.}(2014){Peretto}, {Fuller}, {Andr{\'e}},
  {Arzoumanian}, {Rivilla}, {Bardeau}, {Duarte Puertas}, {Guzman Fernandez},
  {Lenfestey}, {Li}, {Olguin}, {R{\"o}ck}, {de Villiers}, \&
  {Williams}}]{Peretto2014}
{Peretto}, N., {Fuller}, G.~A., {Andr{\'e}}, P., {et~al.} 2014, A{\&}A, 561,
  A83, \dodoi{10.1051/0004-6361/201322172}

\bibitem[{{Pineda} {et~al.}(2020){Pineda}, {Segura-Cox}, {Caselli},
  {Cunningham}, {Zhao}, {Schmiedeke}, {Maureira}, \& {Neri}}]{Pineda2020}
{Pineda}, J.~E., {Segura-Cox}, D., {Caselli}, P., {et~al.} 2020, Nature
  Astronomy, 4, 1158, \dodoi{10.1038/s41550-020-1150-z}

\bibitem[{{Pineda} {et~al.}(2022){Pineda}, {Arzoumanian}, {Andr{\'e}},
  {Friesen}, {Zavagno}, {Clarke}, {Inoue}, {Chen}, {Lee}, {Soler}, \&
  {Kuffmeier}}]{Pineda2022}
{Pineda}, J.~E., {Arzoumanian}, D., {Andr{\'e}}, P., {et~al.} 2022, arXiv
  e-prints, arXiv:2205.03935.
\newblock \doarXiv{2205.03935}

\bibitem[{{Portegies Zwart}(2019)}]{PortegiesZwart2019}
{Portegies Zwart}, S. 2019, \aap, 622, A69, \dodoi{10.1051/0004-6361/201833974}

\bibitem[{{Sahijpal} \& {Goswami}(1998)}]{Sahijpal1998}
{Sahijpal}, S., \& {Goswami}, J.~N. 1998, \apjl, 509, L137,
  \dodoi{10.1086/311778}

\bibitem[{{Samal} {et~al.}(2018){Samal}, {Deharveng}, {Zavagno}, {Anderson},
  {Molinari}, \& {Russeil}}]{Samal2018}
{Samal}, M.~R., {Deharveng}, L., {Zavagno}, A., {et~al.} 2018, \aap, 617, A67,
  \dodoi{10.1051/0004-6361/201833015}

\bibitem[{{Schneider} \& {Elmegreen}(1979)}]{Schneider1979}
{Schneider}, S., \& {Elmegreen}, B.~G. 1979, ApJS, 41, 87,
  \dodoi{10.1086/190609}

\bibitem[{{Scott}(2007)}]{Scott2007}
{Scott}, E. R.~D. 2007, Annual Review of Earth and Planetary Sciences, 35, 577,
  \dodoi{10.1146/annurev.earth.35.031306.140100}

\bibitem[{{Shimajiri} {et~al.}(2019){Shimajiri}, {Andr{\'e}}, {Ntormousi},
  {Men'shchikov}, {Arzoumanian}, \& {Palmeirim}}]{Shimajiri2019ALMA}
{Shimajiri}, Y., {Andr{\'e}}, P., {Ntormousi}, E., {et~al.} 2019, \aap, 632,
  A83, \dodoi{10.1051/0004-6361/201935689}

\bibitem[{{Sieverding} {et~al.}(2020){Sieverding}, {M{\"u}ller}, \&
  {Qian}}]{Sieverding2020}
{Sieverding}, A., {M{\"u}ller}, B., \& {Qian}, Y.~Z. 2020, \apj, 904, 163,
  \dodoi{10.3847/1538-4357/abc61b}

\bibitem[{{Sukhbold} {et~al.}(2016){Sukhbold}, {Ertl}, {Woosley}, {Brown}, \&
  {Janka}}]{Sukhbold2016}
{Sukhbold}, T., {Ertl}, T., {Woosley}, S.~E., {Brown}, J.~M., \& {Janka}, H.~T.
  2016, \apj, 821, 38,
  \dodoi{10.3847/0004-637X/821/1/3810.48550/arXiv.1510.04643}

\bibitem[{{Tachibana} \& {Huss}(2003)}]{Tachibana2003}
{Tachibana}, S., \& {Huss}, G.~R. 2003, \apjl, 588, L41

\bibitem[{{Takigawa} {et~al.}(2008){Takigawa}, {Miki}, {Tachibana}, {Huss},
  {Tominaga}, {Umeda}, \& {Nomoto}}]{Takigawa2008}
{Takigawa}, A., {Miki}, J., {Tachibana}, S., {et~al.} 2008, \apj, 688, 1382

\bibitem[{{Tang} \& {Dauphas}(2015)}]{Tang2015}
{Tang}, H., \& {Dauphas}, N. 2015, \apj, 802, 22

\bibitem[{{Telus} {et~al.}(2018){Telus}, {Huss}, {Nagashima}, {Ogliore}, \&
  {Tachibana}}]{Telus2018}
{Telus}, M., {Huss}, G.~R., {Nagashima}, K., {Ogliore}, R.~C., \& {Tachibana},
  S. 2018, \gca, 221, 342

\bibitem[{{Thieme} {et~al.}(2022){Thieme}, {Lai}, {Lin}, {Cheong}, {Lee},
  {Yen}, {Li}, {Lam}, \& {Zhao}}]{Thieme2022}
{Thieme}, T.~J., {Lai}, S.-P., {Lin}, S.-J., {et~al.} 2022, \apj, 925, 32,
  \dodoi{10.3847/1538-4357/ac382b}

\bibitem[{{Thornton} {et~al.}(1998){Thornton}, {Gaudlitz}, {Janka}, \&
  {Steinmetz}}]{Thornton1998}
{Thornton}, K., {Gaudlitz}, M., {Janka}, H.~T., \& {Steinmetz}, M. 1998, \apj,
  500, 95, \dodoi{10.1086/305704}

\bibitem[{{Trappitsch} {et~al.}(2018){Trappitsch}, {Boehnke}, {Stephan},
  {Telus}, {Savina}, {Pardo}, {Davis}, {Dauphas}, {Pellin}, \&
  {Huss}}]{Trappitsch2018}
{Trappitsch}, R., {Boehnke}, P., {Stephan}, T., {et~al.} 2018, \apjl, 857, L15

\bibitem[{{Tsukamoto} {et~al.}(2022){Tsukamoto}, {Maury}, {Commer{\c{c}}on},
  {Alves}, {Cox}, {Sakai}, {Ray}, {Zhao}, \& {Machida}}]{Tsukamoto2022}
{Tsukamoto}, Y., {Maury}, A., {Commer{\c{c}}on}, B., {et~al.} 2022, arXiv
  e-prints, arXiv:2209.13765, \dodoi{10.48550/arXiv.2209.13765}

\bibitem[{{Valdivia-Mena} {et~al.}(2022){Valdivia-Mena}, {Pineda},
  {Segura-Cox}, {Caselli}, {Neri}, {L{\'o}pez-Sepulcre}, {Cunningham},
  {Bouscasse}, {Semenov}, {Henning}, {Pi{\'e}tu}, {Chapillon}, {Dutrey},
  {Fuente}, {Guilloteau}, {Hsieh}, {Jim{\'e}nez-Serra}, {Marino}, {Maureira},
  {Smirnov-Pinchukov}, {Tafalla}, \& {Zhao}}]{Valdivia-Mena2022}
{Valdivia-Mena}, M.~T., {Pineda}, J.~E., {Segura-Cox}, D.~M., {et~al.} 2022,
  arXiv e-prints, arXiv:2208.01023.
\newblock \doarXiv{2208.01023}

\bibitem[{{Vuissoz} {et~al.}(2004){Vuissoz}, {Meynet}, {Kn{\"o}dlseder},
  {Cervi{\~n}o}, {Schaerer}, {Palacios}, \& {Mowlavi}}]{Vuissoz2004}
{Vuissoz}, C., {Meynet}, G., {Kn{\"o}dlseder}, J., {et~al.} 2004, \nar, 48, 7,
  \dodoi{10.1016/j.newar.2003.11.003}

\bibitem[{{Walch} {et~al.}(2012){Walch}, {Whitworth}, {Bisbas}, {W{\"u}nsch},
  \& {Hubber}}]{Walch2012}
{Walch}, S.~K., {Whitworth}, A.~P., {Bisbas}, T., {W{\"u}nsch}, R., \&
  {Hubber}, D. 2012, \mnras, 427, 625, \dodoi{10.1111/j.1365-2966.2012.21767.x}

\bibitem[{{Ward-Thompson} {et~al.}(1999){Ward-Thompson}, {Motte}, \&
  {Andre}}]{Ward-Thompson1999}
{Ward-Thompson}, D., {Motte}, F., \& {Andre}, P. 1999, MNRAS, 305, 143,
  \dodoi{10.1046/j.1365-8711.1999.02412.x}

\bibitem[{{Ward-Thompson} {et~al.}(1994){Ward-Thompson}, {Scott}, {Hills}, \&
  {Andre}}]{Ward-Thompson1994}
{Ward-Thompson}, D., {Scott}, P.~F., {Hills}, R.~E., \& {Andre}, P. 1994,
  MNRAS, 268, 276

\bibitem[{{Williams} {et~al.}(2018){Williams}, {Peretto}, {Avison},
  {Duarte-Cabral}, \& {Fuller}}]{Williams2018}
{Williams}, G.~M., {Peretto}, N., {Avison}, A., {Duarte-Cabral}, A., \&
  {Fuller}, G.~A. 2018, A{\&}A, 613, A11, \dodoi{10.1051/0004-6361/201731587}

\bibitem[{{Young}(2014)}]{Young2014}
{Young}, E.~D. 2014, Earth and Planetary Science Letters, 392, 16,
  \dodoi{10.1016/j.epsl.2014.02.014}

\bibitem[{{Zavagno} {et~al.}(2020){Zavagno}, {Andr{\'e}}, {Schuller},
  {Peretto}, {Shimajiri}, {Arzoumanian}, {Csengeri}, {Figueira}, {Fuller},
  {K{\"o}nyves}, {Men'shchikov}, {Palmeirim}, {Roussel}, {Russeil},
  {Schneider}, \& {Zhang}}]{Zavagno2020}
{Zavagno}, A., {Andr{\'e}}, P., {Schuller}, F., {et~al.} 2020, \aap, 638, A7,
  \dodoi{10.1051/0004-6361/202037815}

\end{thebibliography}



\end{document}